\documentclass[12pt]{article}

\usepackage{epsfig}
\usepackage{latexsym}

\newcommand{\be}{\begin{equation}}
\newcommand{\ee}{\end{equation}}
\newcommand{\bea}{\begin{eqnarray}}
\newcommand{\eea}{\end{eqnarray}}
\newcommand{\vp}{\varphi}

\def\M{{\cal M}}

\def\su#1{{\rm SU}(#1)}

\def\u#1{{\rm U}(#1)}

\def\gapp{\mathrel{\raise.3ex\hbox{$>$}\mkern-14mu
              \lower0.6ex\hbox{$\sim$}}}
\def\gsim{\gapp}
\def\lapp{\mathrel{\raise.3ex\hbox{$<$}\mkern-14mu
              \lower0.6ex\hbox{$\sim$}}}
\def\lsim{\lapp}
\newcommand{\mpl}{M_p}

\begin{document}

\title{TASI Lectures: Introduction to Cosmology}
 
\author{Mark Trodden$^1$ and Sean M. Carroll$^2$ \\
\\
$^1$Department of Physics \\
Syracuse University \\
Syracuse, NY 13244-1130, USA\\
\\
$^2$Enrico Fermi Institute, Department of Physics,\\
and Center for Cosmological Physics\\
University of Chicago \\
5640 S.~Ellis Avenue, Chicago, IL 60637, USA}

\maketitle

\begin{abstract}
These proceedings summarize lectures that were delivered as part of
the 2002 and 2003 Theoretical Advanced Study Institutes in elementary
particle physics (TASI) at the University of Colorado at Boulder. They
are intended to provide a pedagogical introduction to cosmology aimed
at advanced graduate students in particle physics and string theory.
\end{abstract}

\begin{flushright}
SU-GP-04/1-1
\end{flushright}

\vfill

\newpage

\tableofcontents

\newpage

\section{Introduction}
The last decade has seen an explosive increase in both the volume and
the accuracy of data obtained from cosmological observations.  The
number of techniques available to probe and cross-check these data has
similarly proliferated in recent years.

Theoretical cosmologists have not been slouches during this time,
either. However, it is fair to say that we have not made comparable
progress in connecting the wonderful ideas we have to explain the
early universe to concrete fundamental physics models. One of our
hopes in these lectures is to encourage the dialogue between
cosmology, particle physics, and string theory that will be needed to
develop such a connection.

In this paper, we have combined material from two sets of TASI
lectures (given by SMC in 2002 and MT in 2003).  We have taken the
opportunity to add more detail than was originally presented, as well
as to include some topics that were originally excluded for reasons of
time.  Our intent is to provide a concise introduction to the basics
of modern cosmology as given by the standard ``$\Lambda$CDM'' Big-Bang
model, as well as an overview of topics of current research interest.

In Lecture 1 we present the fundamentals of the standard cosmology,
introducing evidence for homogeneity and isotropy and the
Friedmann-Robertson-Walker models that these make possible.  In
Lecture 2 we consider the actual state of our current universe, which
leads naturally to a discussion of its most surprising and problematic
feature: the existence of dark energy.  In Lecture 3 we consider the
implications of the cosmological solutions obtained in Lecture 1 for
early times in the universe. In particular, we discuss thermodynamics
in the expanding universe, finite-temperature phase transitions, and
baryogenesis. Finally, Lecture 4 contains a discussion of the problems
of the standard cosmology and an introduction to our best-formulated
approach to solving them -- the inflationary universe.

Our review is necessarily superficial, given the large number of 
topics relevant to modern cosmology.  More detail can be found in
several excellent textbooks \cite{kt,lindebook,
peebles,peacock2,ryden,dodelson2,ll}.
Throughout the lectures we have borrowed liberally (and sometimes
verbatim) from earlier reviews of our own 
\cite{carroll1,Trodden:1998ym,Riotto:1999yt,Carroll:1999iy,
Albrecht:2001xp,Akerib:2002ez,sag,carroll2}.

Our metric signature is $-+++$.  We use units in which $\hbar=c=1$, and 
define the reduced Planck mass by 
$M_P \equiv (8\pi G)^{-1/2} \simeq 10^{18}$GeV.

\section{Fundamentals of the Standard Cosmology}

\subsection{Homogeneity and Isotropy:  The Robertson-Walker Metric}
Cosmology as the application of general relativity (GR) to the entire  
universe would seem a hopeless endeavor were it not for a remarkable 
fact -- the universe is spatially homogeneous and isotropic on the 
largest scales. 

``Isotropy'' is the claim that the universe looks the same in
all direction.  Direct evidence comes from the smoothness of the 
temperature of the cosmic microwave background, as we will discuss
later.  ``Homogeneity'' is the claim that the universe looks the same
at every point.  It is harder to test directly, although some evidence
comes from number counts of galaxies.  More traditionally, we may
invoke the ``Copernican principle,'' that we do not live in a special
place in the universe.  Then it follows that, since the universe
appears isotropic around us, it should be isotropic around every
point; and a basic theorem of geometry states that isotropy around
every point implies homogeneity.

We may therefore approximate the universe as a spatially homogeneous
and isotropic three-dimensional space which may expand (or, in
principle, contract) as a function of time.  The metric on such a
spacetime is necessarily of the Robertson-Walker (RW) form, as we now
demonstrate.\footnote{One of the authors has a sentimental attachment
to the following argument, since he learned it in his first cosmology
course~\cite{Stewart}.}

Spatial isotropy implies spherical symmetry. Choosing a point as an 
origin, and using coordinates $(r,\theta,\phi)$ around this point, 
the spatial line element must take the form
\begin{equation}
d\sigma^2=dr^2+f^{2}(r)\left(d\theta^2+\sin^2\theta d\phi^2\right) \ ,
\end{equation}
where $f(r)$ is a real function, which, if the metric is to be
nonsingular at the origin, obeys $f(r)\sim r$ as $r\rightarrow 0$.

\begin{figure}[t]
\centerline{
\psfig{figure=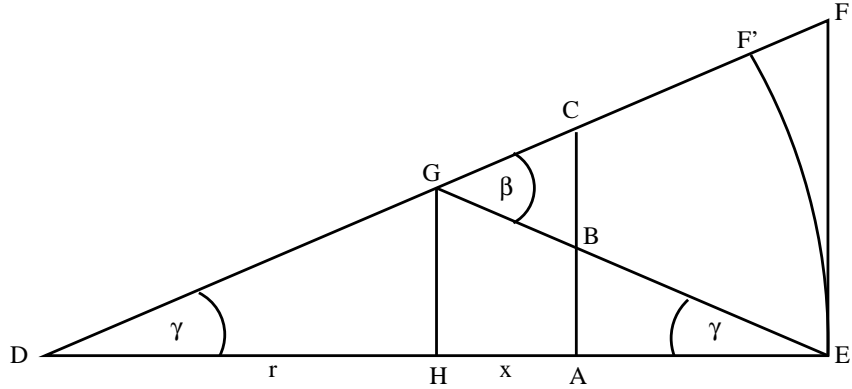,height=2in}}
\caption{Geometry of a homogeneous and isotropic space.}
\label{isotropy}
\end{figure}

Now, consider figure~\ref{isotropy} in the $\theta=\pi/2$ plane. In this 
figure $DH=HE=r$, both $DE$ and $\gamma$ are small and $HA=x$. 
Note that the two angles labeled $\gamma$ are equal because of homogeneity 
and isotropy. Now, note that
\begin{equation}
\label{EFeqn}
EF \simeq EF'=f(2r)\gamma=f(r)\beta \ .
\end{equation}
Also
\begin{equation}
\label{ACeqn}
AC=\gamma f(r+x)=AB+BC=\gamma f(r-x)+\beta f(x) \ .
\end{equation}
Using~(\ref{EFeqn}) to eliminate $\beta/\gamma$, rearranging~(\ref{ACeqn}), 
dividing by $2x$ and taking the limit $x\rightarrow\infty$ yields
\begin{equation}
\label{fde}
\frac{df}{dr}=\frac{f(2r)}{2f(r)} \ .
\end{equation}
We must solve this subject to $f(r)\sim r$ as $r\rightarrow 0$.  It is
easy to check that if $f(r)$ is a solution then $f(r/\alpha)$ is a
solution for constant $\alpha$. Also, $r$, $\sin r$ and $\sinh r$ are
all solutions. Assuming analyticity and writing $f(r)$ as a power
series in $r$ it is then easy to check that, up to scaling, these are
the only three possible solutions.

Therefore, the most general spacetime metric consistent with
homogeneity and isotropy is
\begin{equation}
\label{frwmetric}
ds^2=-dt^2+a^2(t)
\left[ d\rho^2+f^2(\rho)\left(d\theta^2+\sin^2\theta d\phi^2\right)\right] \ ,
\end{equation}
where the three possibilities for $f(\rho)$ are \be f(\rho)=
\{\sin(\rho), ~~ \rho, ~~ \sinh(\rho)\} \ .  \ee 
This is a purely
geometric fact, independent of the details of general relativity.  We
have used spherical polar coordinates $(\rho,\theta,\phi)$, since
spatial isotropy implies spherical symmetry about every point.  The
time coordinate $t$, which is the proper time as measured by a
comoving observer (one at constant spatial coordinates), is referred
to as cosmic time, and the function $a(t)$ is called the scale factor.

There are two other useful forms for the RW metric. First, a simple
change of variables in the radial coordinate yields
\begin{equation}
\label{frwmetric2}
ds^2=-dt^2 +a^2(t)\left[\frac{dr^2}{1-kr^2}
+r^2\left(d\theta^2+\sin^2\theta d\phi^2\right)\right] \ ,
\end{equation}
where
\begin{equation}
\label{curvature}
k=\left\{\begin{array}{ll}
+1 &  \ \ \ \ \ \mbox{if $f(\rho)=\sin(\rho)$} \\
~~0 &  \ \ \ \ \ \mbox{if $f(\rho)=\rho$} \\
-1 &  \ \ \ \ \ \mbox{if $f(\rho)=\sinh(\rho)$} 
\end{array}\right. \ .
\end{equation}

Geometrically, $k$ describes the curvature of the spatial sections
(slices at constant cosmic time). $k=+1$ corresponds to positively
curved spatial sections (locally isometric to 3-spheres); $k=0$
corresponds to local flatness, and $k=-1$ corresponds to negatively
curved (locally hyperbolic) spatial sections. These are all local
statements, which should be expected from a local theory such as
GR. The global topology of the spatial sections may be that of the
covering spaces -- a 3-sphere, an infinite plane or a 3-hyperboloid --
but it need not be, as topological identifications under freely-acting
subgroups of the isometry group of each manifold are allowed. As a
specific example, the $k=0$ spatial geometry could apply just as well
to a 3-torus as to an infinite plane.

Note that we have not chosen a normalization such that $a_0=1$.
We are not free to do this and to simultaneously normalize $|k|=1$,
without including explicit factors of the current scale factor in
the metric.  In the flat case, where $k=0$, we can safely choose
$a_0=1$.

A second change of variables, which may be applied to
either~(\ref{frwmetric}) or~(\ref{frwmetric2}), is to transform to
{\it conformal time}, $\tau$, via
\begin{equation}
\label{conformaltime}
\tau(t)\equiv \int^t \frac{dt'}{a(t')} \ .
\end{equation}
Applying this to~(\ref{frwmetric2}) yields
\begin{equation}
\label{conffrwmetric2}
ds^2= a^2(\tau)\left[-d\tau^2 +
\frac{dr^2}{1-kr^2}+r^2\left(d\theta^2+\sin^2\theta
d\phi^2\right)\right] \ ,
\end{equation}
where we have written $a(\tau)\equiv a[t(\tau)]$ as is conventional.
The conformal time does not measure the proper time for any particular
observer, but it does simplify some calculations.

\begin{figure}
\centerline{
\psfig{figure=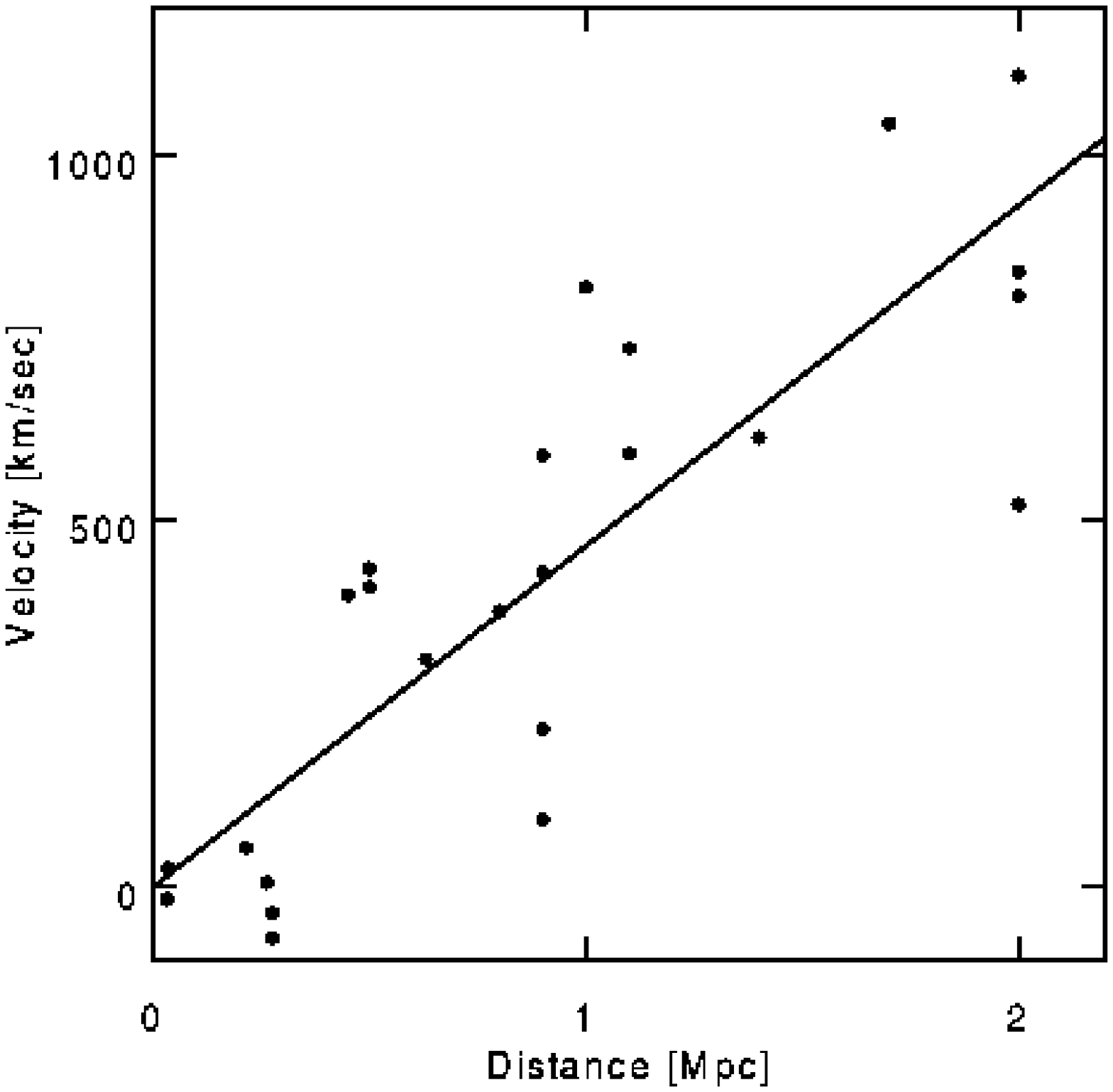,height=3in}
\psfig{figure=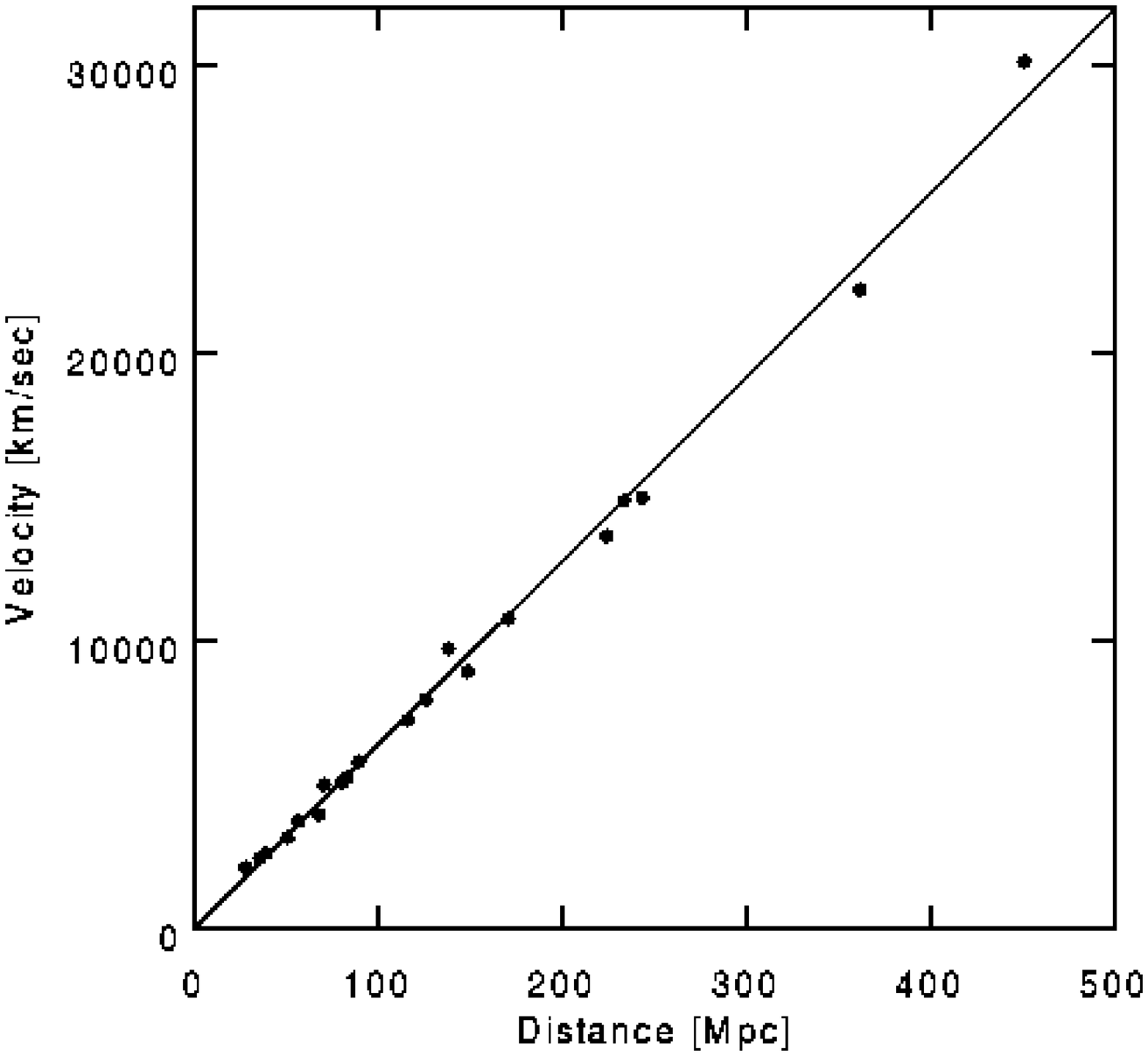,height=3in}}
\caption{Hubble diagrams (as replotted in ~\cite{nwct}) showing the relationship between recessional
velocities of distant galaxies and their distances. The left plot
shows the original data of Hubble~\cite{hubbleoriginal} (and a rather unconvincing
straight-line fit through it). To reassure you, the right plot shows
much more recent data~\cite{Riess:1996pa}, using significantly more distant galaxies (note
difference in scale).}
\label{hubblefig}
\end{figure}

A particularly useful quantity to define from the scale factor is the
{\it Hubble parameter} (sometimes called the Hubble constant), given
by
\begin{equation}
\label{Hubbleconstant}
H\equiv \frac{{\dot a}}{a} \ .
\end{equation}
The Hubble parameter relates how fast the most distant galaxies are
receding from us to their distance from us via Hubble's law, 
\be v \simeq Hd.  
\ee 
This is the relationship that was discovered by Edwin
Hubble, and has been verified to high accuracy by modern observational
methods (see figure~\ref{hubblefig}).

\subsection{Dynamics:  The Friedmann Equations}

As mentioned, the RW metric is a purely kinematic consequence of
requiring homogeneity and isotropy of our spatial sections.  We next
turn to dynamics, in the form of differential equations governing the
evolution of the scale factor $a(t)$.  These will come from applying
Einstein's equation, 
\be R_{\mu\nu}-{1\over 2}Rg_{\mu\nu} =8\pi G
T_{\mu\nu}
  \label{einstein}
\ee
to the RW metric.

Before diving right in, it is useful to consider the types of
energy-momentum tensors $T_{\mu\nu}$ we will typically encounter in
cosmology.  For simplicity, and because it is consistent with much we
have observed about the universe, it is often useful to adopt the
perfect fluid form for the energy-momentum tensor of cosmological
matter. This form is
\begin{equation}
\label{perfectfluid}
T_{\mu\nu} = (\rho + p)U_\mu U_\nu + p g_{\mu\nu}\ ,
\end{equation}
where $U^{\mu}$ is the fluid four-velocity, $\rho$ is the energy
density in the rest frame of the fluid and $p$ is the pressure in that
same frame.  The pressure is necessarily isotropic, for consistency
with the RW metric.  Similarly, fluid elements will be comoving in the
cosmological rest frame, so that the normalized four-velocity in the
coordinates of (\ref{frwmetric2}) will be
\be
  U^\mu = (1,0,0,0)\ .
\ee
The energy-momentum tensor thus takes the form
\be
  T_{\mu\nu} = \left(\matrix{\rho &&&\cr &&&&\cr && p g_{ij} &\cr &&&&}\right)\ ,
\ee
where $g_{ij}$ represents the spatial metric (including the factor of $a^2$).

Armed with this simplified description for matter, we are now ready to
apply Einstein's equation (\ref{einstein}) to
cosmology. Using~(\ref{frwmetric2}) and~(\ref{perfectfluid}), one
obtains two equations. The first is known as the Friedmann equation,
\begin{equation}
\label{Friedmann}
H^2 \equiv \left(\frac{{\dot a}}{a}\right)^2=\frac{8\pi G}{3}\sum_i \rho_i -\frac{k}{a^2} \ ,
\end{equation}
where an overdot denotes a derivative with respect to cosmic time $t$
and $i$ indexes all different possible types of energy in the
universe. This equation is a constraint equation, in the sense that we
are not allowed to freely specify the time derivative $\dot{a}$; it is
determined in terms of the energy density and curvature. The second
equation, which is an evolution equation, is
\begin{equation}
\label{2ndeinsteineqn}
\frac{\ddot a}{a} +\frac{1}{2}\left(\frac{{\dot a}}{a}\right)^2=-4\pi G\sum_i p_i -\frac{k}{2a^2} \ .
\end{equation}
It is often useful to combine~(\ref{Friedmann}) and~(\ref{2ndeinsteineqn}) to obtain the {\it acceleration equation}
\begin{equation}
\label{acceleration}
\frac{{\ddot a}}{a}=-\frac{4\pi G}{3}\sum_i \left(\rho_i +3p_i \right) \ .
\end{equation}
In fact, if we know the magnitudes and evolutions of the different
energy density components $\rho_i$, the Friedmann equation
(\ref{Friedmann}) is sufficient to solve for the evolution uniquely.
The acceleration equation is conceptually useful, but rarely invoked
in calculations.

The Friedmann equation relates the rate of increase of the scale
factor, as encoded by the Hubble parameter, to the total energy
density of all matter in the universe. We may use the Friedmann
equation to define, at any given time, a critical energy density,
\begin{equation}
\label{criticaldensity}
\rho_c\equiv \frac{3H^2}{8\pi G} \ ,
\end{equation}
for which the spatial sections must be precisely flat ($k=0$). 
We then define the density parameter
\begin{equation}
\label{omega}
\Omega_{\rm total} \equiv \frac{\rho}{\rho_c} \ ,
\end{equation}
which allows us to relate the total energy density in the universe to
its local geometry via
\begin{eqnarray}
\Omega_{\rm total}>1 & \Leftrightarrow & k=+1 \nonumber \\
\Omega_{\rm total}=1 & \Leftrightarrow & k=0 \\
\Omega_{\rm total}<1 & \Leftrightarrow & k=-1 \nonumber \ .
\end{eqnarray}
It is often convenient to define the fractions of the critical energy
density in each different component by
\begin{equation}
\Omega_i=\frac{\rho_i}{\rho_c} \ .
\end{equation}

Energy conservation is expressed in GR by the vanishing of the
covariant divergence of the energy-momentum tensor,
\be
  \nabla_\mu T^{\mu\nu} = 0\ .
\ee
Applying this to our assumptions -- the RW metric (\ref{frwmetric2})
and perfect-fluid energy-momentum tensor (\ref{perfectfluid}) --
yields a single energy-conservation equation,
\begin{equation}
\label{energyconservation}
{\dot \rho} + 3H(\rho+p)=0 \ .
\end{equation}
This equation is actually not independent of the Friedmann and
acceleration equations, but is required for consistency.  It implies
that the expansion of the universe (as specified by $H$) can lead to
local changes in the energy density.  Note that there is no notion of
conservation of ``total energy,'' as energy can be interchanged
between matter and the spacetime geometry.

One final piece of information is required before we can think about
solving our cosmological equations: how the pressure and energy
density are related to each other.  Within the fluid approximation
used here, we may assume that the pressure is a single-valued function
of the energy density $p=p(\rho)$. It is often convenient to define an
equation of state parameter, $w$, by
\be
p= w\rho\ .
\ee 
This should be thought of as the instantaneous definition of the 
parameter $w$; it need represent the full equation of state, which would
be required to calculate the behavior of fluctuations.
Nevertheless, many useful cosmological matter sources do
obey this relation with a constant value of $w$. For example, $w=0$ 
corresponds to pressureless matter, or dust -- any collection of
massive non-relativistic particles would qualify.  Similarly,
$w=1/3$ corresponds to a gas of radiation, whether it be actual
photons or other highly relativistic species.

A constant $w$ leads to a great simplification in solving our
equations. In particular, using~(\ref{energyconservation}), we see
that the energy density evolves with the scale factor according to
\begin{equation}
\label{energydensity}
\rho(a) \propto \frac{1}{a(t)^{3(1+w)}} \ .
\end{equation} 
Note that the behaviors of dust ($w=0$) and radiation ($w=1/3$) are
consistent with what we would have obtained by more heuristic
reasoning. Consider a fixed {\it comoving} volume of the universe -
i.e. a volume specified by fixed values of the coordinates, from which
one may obtain the physical volume at a given time $t$ by multiplying
by $a(t)^3$. Given a fixed number of dust particles (of mass $m$)
within this comoving volume, the energy density will then scale just
as the physical volume, i.e. as $a(t)^{-3}$, in agreement
with~(\ref{energydensity}), with $w=0$.

To make a similar argument for radiation, first note that the
expansion of the universe (the increase of $a(t)$ with time) results
in a shift to longer wavelength $\lambda$, or a {\it redshift}, of
photons propagating in this background. A photon emitted with
wavelength $\lambda_e$ at a time $t_e$, at which the scale factor is
$a_e\equiv a(t_e)$ is observed today ($t=t_0$, with scale factor
$a_0\equiv a(t_0)$) at wavelength $\lambda_o$, obeying
\begin{equation}
\label{redshift}
\frac{\lambda_o}{\lambda_e}=\frac{a_0}{a_e}\equiv 1+z \ .
\end{equation}
The redshift $z$ is often used in place of the scale factor.  Because
of the redshift, the energy density in a fixed number of photons in a
fixed comoving volume drops with the physical volume (as for dust) and
by an extra factor of the scale factor as the expansion of the
universe stretches the wavelengths of light. Thus, the energy density
of radiation will scale as $a(t)^{-4}$, once again in agreement
with~(\ref{energydensity}), with $w=1/3$.

Thus far, we have not included a cosmological constant $\Lambda$ in
the gravitational equations. This is because it is equivalent to treat
any cosmological constant as a component of the energy density in the
universe. In fact, adding a cosmological constant $\Lambda$ to
Einstein's equation is equivalent to including an energy-momentum
tensor of the form
\be
  T_{\mu\nu} = -{\Lambda \over 8\pi G} g_{\mu\nu}\ .
\ee
This is simply a perfect fluid with energy-momentum 
tensor~(\ref{perfectfluid}) with
\begin{eqnarray}
\rho_{\Lambda} & = & \frac{\Lambda}{8\pi G} \nonumber \\
p_{\Lambda} & = & -\rho_{\Lambda} \ , 
\end{eqnarray}
so that the equation-of-state parameter is
\be
  w_{\Lambda}=-1\ .
\ee 
This implies that the energy density is constant,
\be
 \rho_\Lambda = {\rm constant}\ .
\ee
Thus, this energy is constant throughout spacetime; we say that
the cosmological constant is equivalent to {\it vacuum
energy}. 

Similarly, it is sometimes useful to think of any nonzero spatial curvature as yet another component of the 
cosmological energy budget, obeying
\begin{eqnarray}
\rho_{\rm curv} & = & -\frac{3k}{8\pi Ga^2} \nonumber \\
p_{\rm curv} & = & \frac{k}{8\pi Ga^2}\ , 
\end{eqnarray}
so that 
\be
w_{\rm curv}=-1/3 \ .
\ee
It is not an energy density, of course; $\rho_{\rm curv}$ is
simply a convenient way to keep track of how much energy density
is {\rm lacking}, in comparison to a flat universe.

\subsection{Flat Universes}
It is much easier to find exact solutions to cosmological equations of motion when $k=0$. Fortunately for us, nowadays we are able to appeal to more than mathematical simplicity to make this choice. Indeed, as we shall see in later lectures, modern cosmological observations, in particular precision measurements of the cosmic microwave background, show the universe today to be extremely spatially flat.

In the case of flat spatial sections and a constant equation of state parameter $w$, we may exactly solve the Friedmann equation 
(\ref{energydensity}) to obtain
\begin{equation}
\label{flatsolution}
a(t)=a_0 \left(t\over t_0\right)^{2/3(1+w)} \ ,
\end{equation}
where $a_0$ is the scale factor today, unless $w=-1$, 
in which case one obtains $a(t) \propto e^{Ht}$.
Applying this result to some of our favorite energy density sources yields table~\ref{sourcestable}.
\begin{table}[t]
\begin{center}
\begin{tabular}{l|l|l}
Type of Energy & $\rho(a)$ & $a(t)$ \\ \hline
Dust & $a^{-3}$ & $t^{2/3}$ \\
Radiation & $a^{-4}$ & $t^{1/2}$ \\
Cosmological Constant & constant & $e^{Ht}$ \\
\end{tabular}
\end{center}
\caption{A summary of the behaviors of the most important sources of energy density in cosmology.  The behavior of the scale factor applies
to the case of a flat universe; the behavior of the energy densities
is perfectly general.}
\label{sourcestable}
\end{table}

Note that the matter- and radiation-dominated flat universes begin
with $a=0$; this is a singularity, known as the Big Bang.  We can
easily calculate the age of such a universe:
\be
  t_0 = \int_0^1 {da \over aH(a)} =
  {2\over 3(1+w)H_0}\ .
\ee
Unless $w$ is close to $-1$, it is often useful to approximate this
answer by
\be
  t_0 \sim H_0^{-1}\ .
\ee
It is for this reason that the quantity $H_0^{-1}$ is known as the
{\it Hubble time}, and provides a useful estimate of the time scale
for which the universe has been around.

\subsection{Including Curvature}
It is true that we know observationally that the universe today is flat to a high degree of accuracy. However, it is instructive, and useful when considering early cosmology, to consider how the solutions we have already identified change when curvature is included. Since we include this mainly for illustration we will focus on the separate cases of dust-filled and radiation-filled FRW models with zero cosmological constant. This calculation is an example of one that is made much easier by working in terms of conformal time $\tau$.

Let us first consider models in which the energy density is dominated
by matter ($w=0$).
In terms of conformal time the Einstein equations become
\begin{eqnarray}
3(k+h^2) & = & 8\pi G\rho a^2 \nonumber \\
k+h^2+2h' & = & 0 \ , 
\end{eqnarray}
where a prime denotes a derivative with respect to conformal time and  $h(\tau) \equiv a'/a$. These equations are then easily solved for $h(\tau)$ giving
\begin{equation}
\label{hoftaudust}
h(\tau)=\left\{\begin{array}{ll}
\cot(\tau/2) & \ \ \ \ k=1 \\
2/\tau & \ \ \ \ k=0 \\
\coth(\tau/2) & \ \ \ \ k=-1
\end{array} \right. \ .
\end{equation}
This then yields
\begin{equation}
\label{aoftaudust}
a(\tau)\propto \left\{\begin{array}{ll}
1-\cos(\tau) & \ \ \ \ k=1 \\
\tau^2/2 & \ \ \ \ k=0 \\
\cosh(\tau)-1 & \ \ \ \ k=-1
\end{array} \right. \ .
\end{equation}

One may use this to derive the connection between cosmic time and conformal time, which here is
\begin{equation}
\label{toftaudust}
t(\tau)\propto \left\{\begin{array}{ll}
\tau-\sin(\tau) & \ \ \ \ k=1 \\
\tau^3/6 & \ \ \ \ k=0 \\
\sinh(\tau)-\tau & \ \ \ \ k=-1
\end{array} \right. \ .
\end{equation}

Next we consider models dominated by radiation ($w=1/3$).
In terms of conformal time the Einstein equations become
\begin{eqnarray}
3(k+h^2) & = & 8\pi G\rho a^2 \nonumber \\
k+h^2+2h' & = & -\frac{8\pi G \rho}{3}a^2 \ .
\end{eqnarray}
Solving as we did above yields
\begin{equation}
\label{hoftaurad}
h(\tau)=\left\{\begin{array}{ll}
\cot(\tau) & \ \ \ \ k=1 \\
1/\tau & \ \ \ \ k=0 \\
\coth(\tau) & \ \ \ \ k=-1
\end{array} \right. \ ,
\end{equation}
\begin{equation}
\label{aoftaurad}
a(\tau)\propto \left\{\begin{array}{ll}
\sin(\tau) & \ \ \ \ k=1 \\
\tau & \ \ \ \ k=0 \\
\sinh(\tau) & \ \ \ \ k=-1
\end{array} \right. \ ,
\end{equation}
and
\begin{equation}
\label{toftaurad}
t(\tau)\propto \left\{\begin{array}{ll}
1-\cos(\tau) & \ \ \ \ k=1 \\
\tau^2/2 & \ \ \ \ k=0 \\
\cosh(\tau)-1 & \ \ \ \ k=-1
\end{array} \right. \ .
\end{equation}

It is straightforward to interpret these solutions by examining the
behavior of the scale factor $a(\tau)$; the qualitative features are
the same for matter- or radiation-domination.  In both cases, the
universes with positive curvature ($k=+1$) expand from an initial
singularity with $a=0$, and later recollapse again.  The initial
singularity is the Big Bang, while the final singularity is sometimes
called the Big Crunch.  The universes
with zero or negative curvature begin at the Big Bang and
expand forever.  This behavior is 
not inevitable, however; we will see below how it can be altered by
the presence of vacuum energy.

\subsection{Horizons}
One of the most crucial concepts to master about FRW models is the existence of {\it horizons}. This concept will prove useful in a variety of places in these lectures, but most importantly in understanding the shortcomings of what we are terming the standard cosmology.

\begin{figure}
\centerline{
\psfig{figure=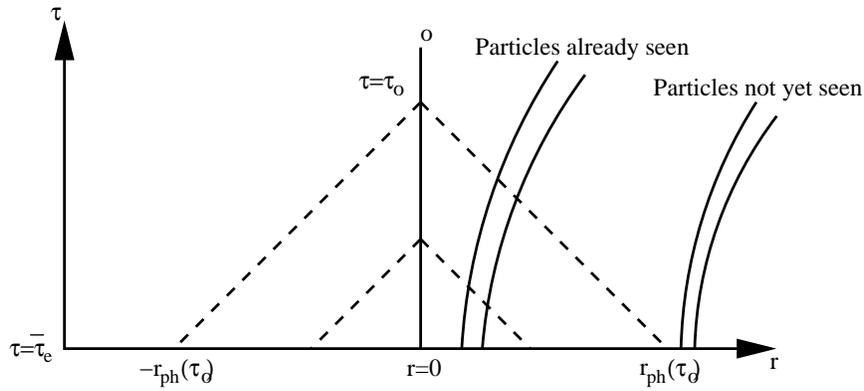,height=2in}}
\caption{Particle horizons arise when the past light cone of an observer
$o$ terminates at a finite conformal time.  Then there will be worldlines
of other particles which do not intersect the past of $o$, meaning that
they were never in causal contact.}
\label{parthorfig}
\end{figure}

Suppose an emitter, $e$, sends a light signal to an observer, $o$, who is at $r=0$. Setting $\theta={\rm constant}$ and $\phi={\rm constant}$ and working in conformal time, for such radial null rays we have $\tau_o -\tau=r$.  In particular this means that
\begin{equation}
\tau_o -\tau_e = r_e \ .
\end{equation}
Now suppose $\tau_e$ is bounded below by ${\bar \tau}_e$; for example,
$\bar\tau_e$ might represent the Big Bang singularity. 
Then there exists a maximum distance to which the observer can see, known as the {\it particle horizon distance}, given by
\begin{equation}
\label{confparthor}
r_{\rm ph}(\tau_o)=\tau_o -{\bar \tau}_e \ .
\end{equation}
The physical meaning of this is illustrated in figure~\ref{parthorfig}.

\begin{figure}
\centerline{
\psfig{figure=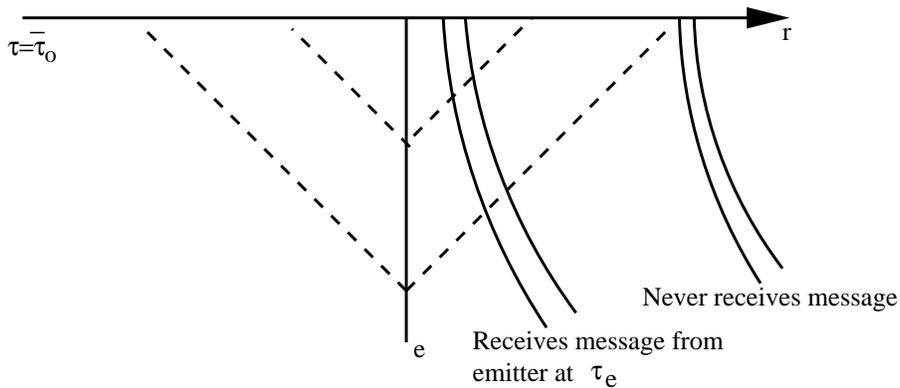,height=2in}}
\caption{Event horizons arise when the future light cone of an observer
$o$ terminates at a finite conformal time.  Then there will be worldlines
of other particles which do not intersect the future of $o$, meaning that
they cannot possibly influence each other.}
\label{eventhorfig}
\end{figure}

Similarly, suppose $\tau_o$ is bounded above by ${\bar \tau}_o$. Then there exists a limit to spacetime events which can be influenced by the emitter. This limit is known as the {\it event horizon distance}, given by
\begin{equation}
\label{confeventthor}
r_{\rm eh}(\tau_o)={\bar \tau}_o -\tau_e \ ,
\end{equation}
with physical meaning illustrated in figure~\ref{eventhorfig}.

These horizon distances may be converted to {\it proper horizon distances} at cosmic time $t$, for example
\begin{equation}
\label{horizon}
d_H\equiv a(\tau)r_{\rm ph}=a(\tau)(\tau-{\bar \tau}_e)=a(t)\int_{t_e}^t \frac{dt'}{a(t')} \ .
\end{equation}
Just as the Hubble time $H_0^{-1}$ provides a rough guide for the age
of the universe, the Hubble distance $cH_0^{-1}$ provides a rough
estimate of the horizon distance in a matter- or radiation-dominated
universe.

\subsection{Geometry, Destiny and Dark Energy}
In subsequent lectures we will use what we have learned here to 
extrapolate back to some of the earliest times in the universe. We will 
discuss the thermodynamics of the early universe, and the resulting 
interdependency between particle physics and cosmology. However, before 
that, we would like to explore some implications for the future of the 
universe.

For a long time in cosmology, it was quite commonplace to refer to the three possible geometries consistent with homogeneity and isotropy as closed ($k=1$), open ($k=-1$) and flat ($k=0$). There were two reasons for this. First, if one considered only the universal covering spaces, then a positively curved universe would be a 3-sphere, which has finite volume and hence is closed, while a negatively curved universe would be the hyperbolic 3-manifold ${\cal H}^3$, which has infinite volume and hence is open.

Second, with dust and radiation as sources of energy density, universes with greater than the critical density would ultimately collapse, while those with less than the critical density would expand forever, with flat universes lying on the border between the two. for the case of pure dust-filled universes this is easily seen from~(\ref{aoftaudust}) and~(\ref{aoftaurad}).

As we have already mentioned, GR is a local theory, so the first of these points was never really valid. For example, there exist perfectly good compact hyperbolic manifolds, of finite volume, which are consistent with all our cosmological assumptions. However, the connection between geometry and destiny implied by the second point above was quite reasonable as long as dust and radiation were the only types of energy density relevant in the late universe.

In recent years it has become clear that the dominant component of energy 
density in the present universe is neither dust nor radiation, but rather 
is dark energy. This component is characterized by an equation of state 
parameter $w<-1/3$. We will have a lot more to say about this component 
(including the observational evidence for it) in the next lecture,
but for now we would just like to focus on the way in which it has 
completely separated our concepts of geometry and destiny.

For simplicity, let's focus on what happens if the only energy density in the universe is a cosmological constant, with $w=-1$. In this case, the Friedmann equation may be solved for any value of the spatial curvature parameter $k$. If $\Lambda >0$ then the solutions are
\begin{equation}
\label{positiveLambdasolns}
\frac{a(t)}{a_0}=\left\{\begin{array}{ll}
\cosh\left(\sqrt{\frac{\Lambda}{3}} t\right) & \ \ \ \ k=+1  \\
\exp\left(\sqrt{\frac{\Lambda}{3}} t\right) & \ \ \ \ k=0  \\
\sinh\left(\sqrt{\frac{\Lambda}{3}} t\right) & \ \ \ \ k=-1 
\end{array}\right. \ ,
\end{equation}
where we have encountered the $k=0$ case earlier. It is immediately clear that, in the $t \rightarrow\infty$ limit, all solutions expand exponentially, independently of the spatial curvature. In fact, these solutions are all exactly the same spacetime - {\it de~Sitter space} - just in different coordinate systems. These features of de Sitter space will resurface crucially when we discuss inflation. However, the point here is that the universe clearly expands forever in these spacetimes, irrespective of the value of the spatial curvature.  Note, however, that not
all of the solutions in (\ref{positiveLambdasolns}) actually cover all
of de~Sitter space; the $k=0$ and $k=-1$ solutions represent coordinate
patches which only cover part of the manifold.

For completeness, let us complete the description of spaces with a cosmological constant by considering the case $\Lambda<0$. This spacetime is called {\it Anti-de Sitter space} (AdS) and it should be clear from the Friedmann equation that such a spacetime can only exist in a space with spatial curvature $k=-1$. The corresponding solution for the scale factor is
\begin{equation}
\label{antidesitter}
a(t)=a_0 \sin\left(\sqrt{-\frac{\Lambda}{3}} t\right) \ .
\end{equation}
Once again, this solution does not cover all of AdS; for a more complete
discussion, see \cite{ads}.

\section{Our Universe Today and Dark Energy}

In the previous lecture we set up the tools required to analyze the
kinematics and dynamics of homogeneous and isotropic cosmologies in
general relativity.  In this lecture we turn to the actual universe
in which we live, and discuss the remarkable properties cosmologists
have discovered in the last ten years.  Most remarkable among them
is the fact that the universe is dominated by a uniformly-distributed
and slowly-varying source of ``dark energy,'' which may be a
vacuum energy (cosmological constant), a dynamical field, or something
even more dramatic.

\subsection{Matter:  Ordinary and Dark}

In the years before we knew that dark energy was an important
constituent of the universe, and before observations of 
galaxy distributions and CMB anisotropies
had revolutionized the study of structure in the universe, observational
cosmology sought to measure two numbers:  the Hubble constant
$H_0$ and the matter density parameter $\Omega_{\rm M}$.  Both of
these quantities remain undeniably important, even though we have
greatly broadened the scope of what we hope to measure.  The Hubble
constant is often parameterized in terms of a dimensionless
quantity $h$ as
\be
  H_0= 100h {\rm ~km/sec/Mpc}\ .
\ee
After years of effort, determinations of this number seem to have
zeroed in on a largely agreed-upon value;
the Hubble Space Telescope Key Project on the extragalactic
distance scale \cite{mould} finds 
\be
  h = 0.71 \pm 0.06\ ,
  \label{hstkey}
\ee
which is consistent with other methods \cite{freedman}, and what
we will assume henceforth.

For years, determinations of $\Omega_{\rm M}$ based on dynamics
of galaxies and clusters have yielded values between approximately
$0.1$ and $0.4$, noticeably smaller than the critical density.
The last several years have witnessed a number of new methods
being brought to bear on the question; here we sketch some of
the most important ones.

The traditional method to estimate
the mass density of the universe is to ``weigh'' a cluster of
galaxies, divide by its luminosity, and extrapolate the
result to the universe as a whole.  Although clusters are not
representative samples of the universe, they are sufficiently
large that such a procedure has a chance of working.  Studies
applying the virial theorem to cluster dynamics have typically
obtained values $\Omega_{\rm M} = 0.2 \pm 0.1$
\cite{carlberg96,Dekel:1996,Bahcall:1998}.  
Although it is possible that
the global value of $M/L$ differs appreciably from its value
in clusters, extrapolations from small scales do not seem
to reach the critical density \cite{Bahcall:1995}.  New
techniques to weigh the clusters, including gravitational
lensing of background galaxies \cite{smail} and temperature
profiles of the X-ray gas \cite{lewis}, while not yet in
perfect agreement with each other, reach essentially 
similar conclusions.

Rather than measuring the mass relative to the luminosity
density, which may be different inside and outside clusters,
we can also measure it with respect to the baryon density
\cite{white},
which is very likely to have the same value in clusters as
elsewhere in the universe, simply because there is no way
to segregate the baryons from the dark matter on such large
scales.  Most of the baryonic mass is in the hot intracluster
gas \cite{Fukugita:1997bi}, and the fraction $f_{\rm gas}$ 
of total mass in this form can be measured either by
direct observation of X-rays from the gas 
\cite{Mohr:1999} or by distortions
of the microwave background by scattering off hot electrons
(the Sunyaev-Zeldovich effect) \cite{carlstrom}, typically
yielding $0.1 \leq f_{\rm gas} \leq 0.2$.
Since primordial nucleosynthesis provides a determination
of $\Omega_{\rm B}\sim 0.04$, these measurements imply
\begin{equation}
  \Omega_{\rm M} = \Omega_{\rm B}/f_{\rm gas}
  = 0.3\pm 0.1 \ ,
\end{equation}
consistent with the value determined from mass to light ratios.

Another handle on the density parameter in matter comes from
properties of clusters at high redshift.  The very existence
of massive clusters has been used to argue in favor of 
$\Omega_{\rm M}\sim 0.2$ \cite{bf2}, and the lack of appreciable 
evolution of clusters from high redshifts to the present 
\cite{bfc,carlberg97} provides additional evidence that
$\Omega_{\rm M} < 1.0$.  On the other hand, a recent measurement
of the relationship between the temperature and luminosity of
X-ray clusters measured with the XMM-Newton satellite 
\cite{vauclair} has been interpreted as evidence for
$\Omega_{\rm M}$ near unity.  This last result seems at odds with a
variety of other determinations, so we should keep a careful watch
for further developments in this kind of study.

The story of large-scale motions is more ambiguous.  The
peculiar velocities of galaxies are sensitive to the underlying
mass density, and thus to $\Omega_{\rm M}$, but also to the
``bias'' describing the relative amplitude of fluctuations in
galaxies and mass \cite{Dekel:1996,dekel97}.  Nevertheless,
recent advances in very large redshift surveys have led to
relatively firm determinations of the mass density; the 2df
survey, for example, finds 
$0.1 \leq \Omega_{\rm M} \leq 0.4$ \cite{Verde:2001sf}.

Finally, the matter density parameter can be extracted from
measurements of the power spectrum of density fluctuations
(see for example \cite{peacock}).  
As with the CMB, predicting
the power spectrum requires both an assumption of the correct
theory and a specification of a number of cosmological
parameters.  In simple models ({\it e.g.}, with only cold dark
matter and baryons, no massive neutrinos), the spectrum can be
fit (once the amplitude is normalized) by a single ``shape
parameter'', which is found to be equal to $\Gamma =
\Omega_{\rm M}h$.  (For more complicated models see
\cite{eh}.)  Observations then yield $\Gamma \sim 0.25$,
or $\Omega_{\rm M}\sim 0.36$.  For a more careful comparison
between models and observations, see 
\cite{Liddle:1993fq,Liddle:1996,dodelson,primack}.

Thus, we have a remarkable convergence on values for the
density parameter in matter:
\begin{equation}
  0.1 \leq \Omega_{\rm M} \leq 0.4\ .
\end{equation}
As we will see below, this value is in excellent agreement with
that which we would determine indirectly from combinations of
other measurements.

As you are undoubtedly aware, however, matter comes in different
forms; the matter we infer from its gravitational influence
need not be the same kind of ordinary matter we are familiar with from
our experience on Earth.  By ``ordinary matter'' we mean anything
made from atoms and their constituents (protons, neutrons, 
and electrons); this would include all of
the stars, planets, gas and dust in the universe, immediately visible or
otherwise.  Occasionally such matter is referred to as 
``baryonic matter'', where ``baryons'' include
protons, neutrons, and related particles (strongly interacting
particles carrying a conserved quantum number known as ``baryon
number'').  Of course electrons are
conceptually an important part of ordinary matter, but by mass they
are negligible compared to protons and neutrons; 
the mass of ordinary matter comes overwhelmingly from baryons.

Ordinary baryonic matter, it turns out, is not nearly enough to
account for the observed matter density.
Our current best estimates for the baryon density \cite{schramm,
burles} yield
\be
  \Omega_{\rm b} = 0.04 \pm 0.02\ ,
  \label{8.146}
\ee
where these error bars are conservative by most standards.  
This determination comes from a variety of methods:  direct counting of
baryons (the least precise method), consistency with the CMB power
spectrum (discussed later in this lecture), 
and agreement with the predictions of
the abundances of light elements for Big-Bang nucleosynthesis 
(discussed in the next lecture).  Most of the matter density must 
therefore be in the form of non-baryonic
dark matter, which we will abbreviate to simply ``dark matter''.
(Baryons can be dark, but it is increasingly common to reserve the
terminology for the non-baryonic component.)  
Essentially every known particle in the Standard Model
of particle physics has been ruled out as a candidate for this
dark matter.  One of the few things
we know about the dark matter is that is must be ``cold'' --- not
only is it non-relativistic today, but it must have been that
way for a very long time.  If the dark matter were ``hot'', it would
have free-streamed out of overdense regions, suppressing the
formation of galaxies.  The other thing we know about cold dark
matter (CDM) is that it should interact very weakly with ordinary
matter, so as to have escaped detection thus far.  In the next
lecture we will discuss some currently popular candidates for
cold dark matter.

\subsection{Supernovae and the Accelerating Universe}

The great story of {\it fin de siecle} cosmology was the discovery
that matter does not dominate the universe; we need some form of
dark energy to explain a variety of observations.  The first direct
evidence for this finding came from studies using Type Ia supernovae as ``standardizable candles,'' which we now examine.  For more detailed
discussion of both the observational situation and the attendant
theoretical problems, see \cite{weinberg,sahni,carroll1,peeblesratra,
paddy,carroll2}.

Supernovae are rare ---
perhaps a few per century in a Milky-Way-sized galaxy --- but 
modern telescopes allow observers to probe very deeply into 
small regions of the sky, covering a very large number of galaxies
in a single observing run.  Supernovae are also bright, and Type Ia's
in particular all seem to be of nearly uniform intrinsic luminosity
(absolute magnitude $M\sim -19.5$, typically comparable to the
brightness of the entire host galaxy in which they appear)
\cite{branch}.  They
can therefore be detected at high redshifts ($z\sim 1$), 
allowing in principle a good handle on cosmological effects
\cite{tammann79,Goobar:1995}.  

The fact that all SNe Ia are of similar intrinsic luminosities fits 
well with our understanding of these events as explosions which occur
when a white dwarf, onto which mass is gradually accreting from
a companion star, crosses the Chandrasekhar limit and explodes.
(It should be noted that our understanding of supernova
explosions is in a state of development, and theoretical models
are not yet able to accurately reproduce all of the important
features of the observed events.  See \cite{woosley,hachisu,hoflich}
for some recent work.)
The Chandrasekhar limit is a nearly-universal quantity, so it is
not a surprise that the resulting explosions are of nearly-constant
luminosity.  However, there is still a scatter of approximately
$40\% $ in the peak brightness observed in nearby supernovae, which
can presumably be traced to differences in the composition of the
white dwarf atmospheres.  Even if we could collect enough data
that statistical errors could be reduced to a minimum, the existence
of such an uncertainty would cast doubt on any attempts to study
cosmology using SNe~Ia as standard candles.  

Fortunately, the observed differences
in peak luminosities of SNe~Ia are very closely correlated with
observed differences in the shapes of their light curves:
dimmer SNe decline more rapidly after maximum brightness, while
brighter SNe decline more slowly \cite{phillips,rpk,hamuy}.
There is thus a one-parameter family of events, and measuring the
behavior of the light curve along with the apparent luminosity
allows us to largely correct for the intrinsic differences in
brightness, reducing the scatter from $40\% $ to less than $15\% $ 
--- sufficient precision to distinguish between cosmological models.
(It seems likely that the single parameter can be traced to the
amount of $^{56}$Ni produced in the supernova explosion; more
nickel implies both a higher peak luminosity and a higher 
temperature and thus opacity, leading to a slower decline.
It would be an exaggeration, however, to claim that this behavior
is well-understood theoretically.)

\begin{figure}[t]
\centerline{
\psfig{figure=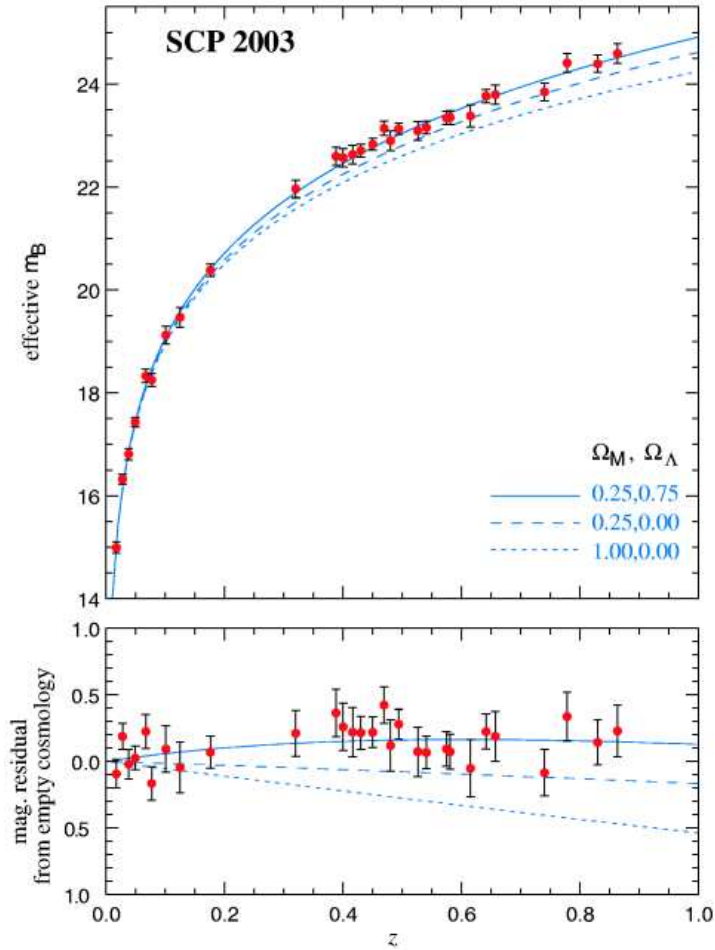,height=5in}}
\caption{Hubble diagram from the Supernova Cosmology Project, as of 2003
\cite{knop}.}
\label{snhubble}
\end{figure}

Following pioneering work reported in \cite{norgard},
two independent groups undertook searches for distant
supernovae in order to measure cosmological parameters:
the High-Z Supernova Team 
\cite{garnavich1,schmidt,riess1,garnavich2,Tonry:2003zg}, 
and the Supernova Cosmology Project
\cite{perlmutter1,perlmutter2,perlmutter3,knop}.
A plot of redshift vs.\ corrected apparent magnitude from the
original SCP data is shown in Figure \ref{snhubble}.
The data are much better fit by a universe dominated by a cosmological
constant than by a flat matter-dominated model.  In fact the
supernova results alone allow a substantial range of possible
values of $\Omega_{\rm M}$ and $\Omega_\Lambda$; however, if we
think we know something about one of these parameters, the other
will be tightly constrained.  In particular, if $\Omega_{\rm M}
\sim 0.3$, we obtain
\be
  \Omega_\Lambda \sim 0.7 \ .
\ee
This corresponds to a vacuum energy density
\be
  \rho_\Lambda \sim 10^{-8} {\rm ~erg/cm^3} \sim (10^{-3}{\rm ~eV})^4\ .
  \label{rhovacobs}
\ee
Thus, the supernova studies have provided direct evidence for
a nonzero value for Einstein's cosmological constant.

Given the significance of these results, it is natural to
ask what level of confidence we should have in them.
There are a number of potential sources of systematic
error which have been considered by the two teams; see
the original papers \cite{schmidt,riess1,perlmutter3}
for a thorough discussion.  Most impressively, the universe implied
by combining the supernova results with direct determinations of
the matter density is spectacularly confirmed by measurements of
the cosmic microwave background, as we discuss in the next section.
Needless to say, however, it would be very useful to have a better
understanding of both the theoretical basis for Type Ia luminosities,
and experimental constraints on possible systematic errors.  
Future experiments, including a proposed
satellite dedicated to supernova cosmology \cite{snap}, will both
help us improve our understanding of the physics of supernovae
and allow a determination of the distance/redshift relation to
sufficient precision to distinguish between the effects of a
cosmological constant and those of more mundane astrophysical
phenomena.  

\subsection{The Cosmic Microwave Background}

Most of the radiation we observe in the universe today is in the form
of an almost isotropic blackbody spectrum, with temperature
approximately 2.7K, known as the {\it Cosmic Microwave Background}
(CMB).  The small angular fluctuations in temperature of the CMB
reveal a great deal about the constituents of the universe, as we now
discuss.

We have mentioned several times the way in which a radiation gas
evolves in and sources the evolution of an expanding FRW universe. It
should be clear from the differing evolution laws for radiation and
dust that as one considers earlier and earlier times in the universe,
with smaller and smaller scale factors, the ratio of the energy
density in radiation to that in matter grows proportionally to
$1/a(t)$.  Furthermore, even particles which are now massive and
contribute to matter used to be hotter, and at sufficiently early
times were relativistic, and thus contributed to radiation.
Therefore, the early universe was dominated by radiation.

At early times the CMB photons were easily energetic enough to ionize
hydrogen atoms and therefore the universe was filled with a charged
plasma (and hence was opaque).  This phase lasted until the photons
redshifted enough to allow protons and electrons to combine, during
the era of {\it recombination}. Shortly after this time, the photons
decoupled from the now-neutral plasma and free-streamed through the
universe.

In fact, the concept of an expanding universe provides us with a clear
explanation of the origin of the CMB. Blackbody radiation is emitted
by bodies in thermal equilibrium. The present universe is certainly
not in this state, and so without an evolving spacetime we would have
no explanation for the origin of this radiation. However, at early
times, the density and energy densities in the universe were high
enough that matter was in approximate thermal equilibrium at each
point in space, yielding a blackbody spectrum at early times.

We will have more to say about thermodynamics in the expanding
universe in our next lecture. However, we should point out one crucial
thermodynamic fact about the CMB. A blackbody distribution, such as
that generated in the early universe, is such that at temperature $T$,
the energy flux in the frequency range $[\nu,\nu+d\nu]$ is given by
the Planck distribution
\begin{equation}
\label{planckdist}
P(\nu,T)d\nu =8\pi h\left(\frac{\nu}{c}\right)^3 \frac{1}{e^{h\nu/kT} -1}d\nu \ ,
\end{equation}
where $h$ is Planck's constant and $k$ is the Boltzmann constant.
Under a rescaling $\nu\rightarrow\alpha\nu$, with $\alpha$=constant,
the shape of the spectrum is unaltered if $T\rightarrow T/\alpha$. We
have already seen that wavelengths are stretched with the cosmic
expansion, and therefore that frequencies will scale inversely due to
the same effect.  We therefore conclude that the effect of cosmic
expansion on an initial blackbody spectrum is to retain its blackbody
nature, but just at lower and lower temperatures,
\be
 T \propto 1/a\ .  
\ee 
This is what we mean when we refer to the
  universe cooling as it expands.  (Note that this strict scaling may
  be altered if energy is dumped into the radiation background during
  a phase transition, as we discuss in the next lecture.)

The CMB is not a perfectly isotropic radiation bath. Deviations from
isotropy at the level of one part in $10^5$ have developed over the
last decade into one of our premier precision observational tools in
cosmology. The small temperature anisotropies on the sky are usually
analyzed by decomposing the signal into spherical harmonics via
\begin{equation}
\label{cmbflucts}
\frac{\Delta T}{T} =\sum_{l,m}a_{lm}Y_{lm}(\theta,\phi) \ ,
\end{equation}
where $a_{lm}$ are expansion coefficients and $\theta$ and $\phi$ are spherical polar angles on the sky.
Defining the power spectrum by
\begin{equation}
\label{cl}
C_l =\langle|a_{lm}|^2\rangle \ ,
\end{equation}
it is conventional to plot the quantity $l(l+1)C_l$ against $l$ in a famous plot that is usually referred to as the CMB power spectrum. 
\begin{figure}
 \centerline{
 \psfig{figure=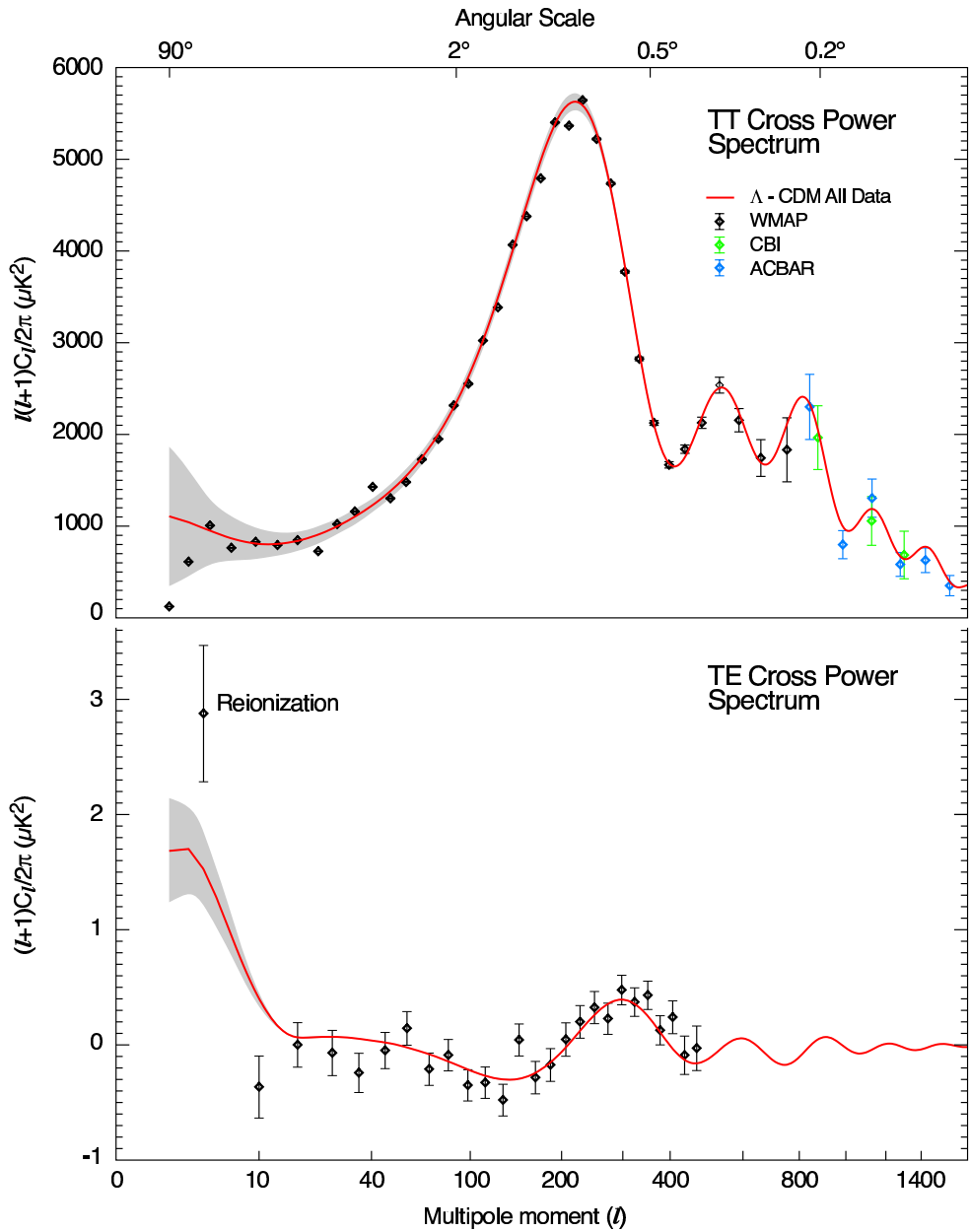,height=6.5in}}
 \caption{The CMB power spectrum from the WMAP satellite \cite{hinshaw}. 
 The error bars on this plot are 1-$\sigma$ and the solid 
 line represents  the best-fit cosmological model \cite{Spergel:2003cb}.
 Also shown is the correlation between the temperature anisotropies
 and the ($E$-mode) polarization.}
 \label{wmappower}
\end{figure}
An example is shown in figure~(\ref{wmappower}), which shows the measurements of the CMB anisotropy from the recent WMAP satellite, as well as a theoretical model (solid line) that fits the data rather well.

These fluctuations in the microwave background are useful to
cosmologists for many reasons. To understand why, we must comment
briefly on why they occur in the first place. Matter today in the
universe is clustered into stars, galaxies, clusters and superclusters
of galaxies. Our understanding of how large scale structure developed
is that initially small density perturbations in our otherwise
homogeneous universe grew through gravitational instability into the
objects we observe today.  Such a picture requires that from place to
place there were small variations in the density of matter at the time
that the CMB first decoupled from the photon-baryon plasma. Subsequent
to this epoch, CMB photons propagated freely through the universe,
nearly unaffected by anything except the cosmic expansion itself.

However, at the time of their decoupling, different photons were
released from regions of space with slightly different gravitational
potentials. Since photons redshift as they climb out of gravitational
potentials, photons from some regions redshift slightly more than
those from other regions, giving rise to a small temperature
anisotropy in the CMB observed today. On smaller scales, the evolution
of the plasma has led to intrinsic differences in the temperature from
point to point.  In this sense the CMB carries with it a fingerprint
of the initial conditions that ultimately gave rise to structure in
the universe.

One very important piece of data that the CMB fluctuations give us is
the value of $\Omega_{\rm total}$. Consider an overdense region of
size $R$, which therefore contracts under self-gravity over a
timescale $R$ (recall $c=1$).  If $R\gg H_{\rm CMB}^{-1}$ then the
region will not have had time to collapse over the lifetime of the
universe at last scattering. If $R\ll H_{\rm CMB}^{-1}$ then collapse
will be well underway at last scattering, matter will have had time to
fall into the resulting potential well and cause a resulting rise in
temperature which, in turn, gives rise to a restoring force from
photon pressure, which acts to damps out the inhomogeneity.

Clearly, therefore, the maximum anisotropy will be on a scale which
has had just enough time to collapse, but not had enough time to
equilibrate - $R\sim H_{\rm CMB}^{-1}$. This means that we expect to
see a peak in the CMB power spectrum at an angular size corresponding
to the horizon size at last scattering. Since we know the physical
size of the horizon at last scattering, this provides us with a ruler
on the sky. The corresponding angular scale will then depend on the
spatial geometry of the universe. For a flat universe ($k=0$,
$\Omega_{\rm total} =1$) we expect a peak at $l\simeq 220$ and, as can
be seen in figure~(\ref{wmappower}), this is in excellent agreement
with observations.

\begin{figure}[t]
\centerline{
\psfig{figure=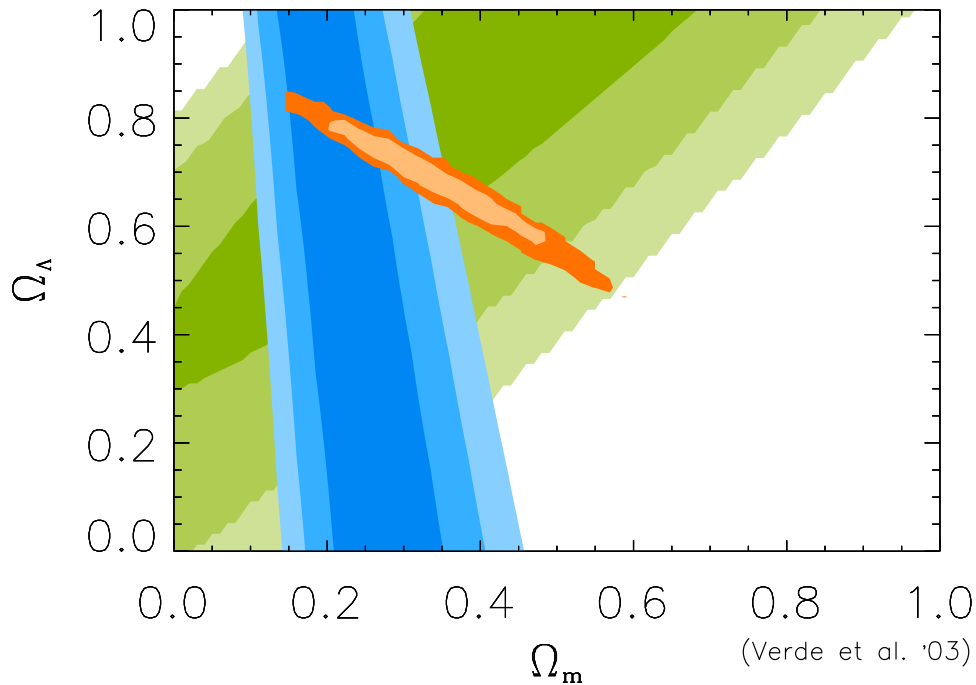,height=4in}}
\caption{Observational constraints in the $\Omega_{\rm M}$-$\Omega_\Lambda$
  plane.  The wide green contours represent constraints from supernovae, 
  the vertical blue contours represent constraints from the 2dF galaxy
  survey, and the small orange contours represent constraints from
  WMAP observations of CMB anisotropies when a prior on the Hubble
  parameter is included.  Courtesy of Licia Verde; see
  \cite{verde} for details.}
\label{combinedplot}
\end{figure}

Beyond this simple heuristic description, careful analysis of all
of the features of the CMB power spectrum (the positions and heights of
each peak and trough) provide constraints on essentially all of the
cosmological parameters.  As an example we consider the results from
WMAP \cite{Spergel:2003cb}.  For the total density of the universe they find
\be
  0.98 \leq \Omega_{\rm total} \leq 1.08
\ee
at $95\%$ confidence -- as mentioned, strong evidence for a flat universe.
Nevertheless, there is still some degeneracy in the parameters, and 
much tighter constraints on the remaining values can be derived by
assuming either an exactly flat universe, or a reasonable value of
the Hubble constant.  When for example we assume a flat universe,
we can derive values for the Hubble constant, matter density (which
then implies the vacuum energy density), and baryon density:
\begin{eqnarray}
h &=& 0.72 \pm 0.05 \nonumber\\
\Omega_{\rm M} = 1- \Omega_\Lambda & = &  0.29 \pm 0.07 \nonumber\\
\Omega_{\rm B} &=& 0.047 \pm 0.006  \nonumber \ .
\end{eqnarray}
If we instead assume that the Hubble constant is given by the
value determined by the HST key project (\ref{hstkey}), we can
derive separate tight constraints on $\Omega_{\rm M}$ and 
$\Omega_\Lambda$; these are shown graphically in 
Figure~\ref{combinedplot}, along with constraints from the supernova
experiments.

Taking all of the data together, we obtain a remarkably consistent
picture of the current constituents of our universe:
\bea
  \Omega_{\rm B} &=& 0.04\cr
  \Omega_{\rm DM} &=& 0.26\cr
  \Omega_\Lambda &=& 0.7\ .
  \label{bestfit}
\eea
Our sense of accomplishment at having measured these numbers is
substantial, although it is somewhat tempered by the realization
that we don't understand any of them.  The
baryon density is mysterious due to the asymmetry between baryons
and antibaryons; as far as dark matter goes, of course, we have
never detected it directly and only have promising ideas as to what
it might be.  Both of these issues will be discussed in the next
lecture.  The biggest mystery is the vacuum energy; we now turn to
an exploration of why it is mysterious and what kinds of mechanisms
might be responsible for its value.

\subsection{The Cosmological Constant Problem(s)}

In classical general relativity the 
cosmological constant $\Lambda$ is 
a completely free parameter.  It has dimensions of [length]$^{-2}$
(while the energy density $\rho_\Lambda$ has units [energy/volume]),
and hence defines a scale, while general
relativity is otherwise scale-free.  Indeed, from purely classical
considerations, we can't even say whether
a specific value of $\Lambda$ is ``large'' or ``small''; it is simply
a constant of nature we should go out and determine through
experiment. 

The introduction of quantum mechanics changes this story somewhat.
For one thing, Planck's constant allows us to define 
the reduced Planck mass $\mpl \sim 10^{18}~{\rm GeV}$, as well as
the reduced Planck length
\be
  L_{\rm P} = \left({8\pi G}\right)^{1/2}
  \sim 10^{-32}~{\rm cm}\ .
\ee
Hence, there is a natural expectation for the scale of the cosmological
constant, namely
\be
  \Lambda^{\rm (guess)} \sim L_{\rm P}^{-2}\ ,
\ee
or, phrased as an energy density,
\be
  \rho_{\rm vac}^{\rm (guess)} \sim M_{\rm P}^4 
  \sim (10^{18}{\rm ~GeV})^4 \sim 10^{112} {\rm ~erg/cm}^3\ .
  \label{rhoguess}
\ee

We can partially justify this guess by thinking about 
quantum fluctuations in the vacuum.  At all energies probed by 
experiment to date, the world is accurately described as a set
of quantum fields (at higher energies it may become strings or
something else).  If we take the Fourier transform
of a free quantum field, each mode of fixed wavelength 
behaves like a simple
harmonic oscillator.  (``Free'' means ``noninteracting''; for our
purposes this is a very good approximation.)  As we know from 
elementary quantum mechanics, the ground-state or zero-point energy
of an harmonic oscillator with potential $V(x)={1\over 2}\omega^2 x^2$
is $E_0 = {1\over 2}\hbar \omega$.  Thus, each mode of a quantum
field contributes to the vacuum energy, and the net result should be
an integral over all of the modes.  Unfortunately this integral
diverges, so the vacuum energy appears to be infinite.
However, the infinity arises from the contribution of modes with
very small wavelengths; perhaps it was a mistake to include such modes,
since we don't really know what might happen at such scales.
To account for our ignorance, we could introduce a cutoff energy,
above which ignore any potential contributions, and
hope that a more complete theory will eventually provide a physical 
justification for doing so.  If this cutoff is at the Planck scale,
we recover the estimate (\ref{rhoguess}).

The strategy of decomposing a free field into individual modes and
assigning a zero-point energy to each one really only makes sense
in a flat spacetime background.  In curved spacetime we can still
``renormalize'' the vacuum energy, relating the classical parameter
to the quantum value by an infinite constant.  After renormalization,
the vacuum energy is completely arbitrary, just as it was in the original
classical theory.  But when we use general relativity we are really
using an effective field theory to describe a certain limit of quantum
gravity.  In the context of effective field theory, if a parameter
has dimensions [mass]$^n$, we expect the corresponding mass parameter
to be driven up to the scale at which the effective description
breaks down.  Hence, if we believe classical
general relativity up to the Planck scale, we would expect the
vacuum energy to be given by our original guess (\ref{rhoguess}).

However, we claim to have measured the vacuum energy (\ref{rhovacobs}).
The observed value is somewhat discrepant with our theoretical estimate:
\be
  \rho_{\rm vac}^{\rm (obs)}
  \sim  10^{-120}\rho_{\rm vac}^{\rm (guess)}\ . 
  \label{rhoobs}
\ee
This is the famous 120-orders-of-magnitude discrepancy that makes
the cosmological constant problem such a glaring embarrassment.
Of course, it is a little unfair to emphasize the factor of
$10^{120}$, which depends on the fact that energy density
has units of [energy]$^4$.  We can express the vacuum energy in
terms of a mass scale,
\be
  \rho_{\rm vac} = M_{\rm vac}^4\ ,
\ee
so our observational result is
\be
  M_{\rm vac}^{{\rm (obs)}} \sim 10^{-3}{\rm ~eV}\ .
\ee
The discrepancy is thus
\be
  M_{\rm vac}^{{\rm (obs)}} \sim 10^{-30} M_{\rm vac}^{{\rm (guess)}}\ .
  \label{mvac1}
\ee
We should think of the cosmological constant problem as a
discrepancy of 30 orders of magnitude in energy scale.  

In addition to the fact that it is very small compared to its natural
value, the vacuum energy presents an additional puzzle:
the coincidence between the
observed vacuum energy and the current matter
density.  Our best-fit universe (\ref{bestfit}) features vacuum
and matter densities of the same order of magnitude, but 
the ratio of these quantities changes rapidly as the universe expands:
\be
  {\Omega_\Lambda \over \Omega_{\rm M}} = {\rho_\Lambda
  \over \rho_{\rm M}} \propto a^3\ .
\ee
As a consequence,
at early times the vacuum energy was negligible in comparison to
matter and radiation, while at late times matter and radiation are
negligible.  There is only a brief epoch of the universe's history
during which it would be possible to
witness the transition from domination by
one type of component to another. 

To date, there are not any especially promising approaches to
calculating the vacuum energy and getting the right answer;
it is nevertheless instructive to consider the example of
supersymmetry, which relates to the cosmological constant problem
in an interesting way.
Supersymmetry posits that for each fermionic degree of freedom 
there is a matching bosonic degree of freedom, and vice-versa.
By ``matching'' we mean, for example, that the spin-1/2 electron
must be accompanied by a spin-0 ``selectron'' with the same mass
and charge.  The good news is that, while bosonic
fields contribute a positive vacuum energy, for fermions the
contribution is negative.  Hence, if degrees of freedom exactly
match, the net vacuum energy sums to zero.  Supersymmetry is thus
an example of a theory, other than gravity, where the absolute
zero-point of energy is a meaningful concept.  (This can be traced
to the fact that supersymmetry is a spacetime symmetry, relating
particles of different spins.)

We do not, however, live in a supersymmetric state; there is no
selectron with the same mass and charge as an electron, or we would
have noticed it long ago.  If supersymmetry exists in nature, it must
be broken at some scale $M_{\rm susy}$.  In a theory with broken supersymmetry,
the vacuum energy is not expected to vanish, but to be of order
\be
  M_{\rm vac} \sim M_{\rm susy}\ ,\qquad \qquad \quad{\rm (theory)}
\ee
with $\rho_{\rm vac} = M_{\rm vac}^4$.  What should $M_{\rm susy}$ be?  One nice
feature of supersymmetry is that it helps us understand the 
hierarchy problem -- why the scale of electroweak symmetry breaking
is so much smaller than the scales of quantum gravity or grand
unification.  For supersymmetry to be relevant to the hierarchy
problem, we need the 
supersymmetry-breaking scale to be just above the electroweak 
scale, or
\be
  M_{\rm susy} \sim 10^3{\rm ~GeV}\ .
\ee
In fact, this is very close to the experimental bound, and there
is good reason to believe that supersymmetry will be discovered 
soon at Fermilab or CERN, if it is connected to electroweak
physics.  

Unfortunately, we are left with a sizable discrepancy between
theory and observation:
\be
  M_{\rm vac}^{\rm (obs)} \sim 10^{-15}M_{\rm susy}\ . \qquad {\rm (experiment)}
  \label{mvac2}
\ee
Compared to (\ref{mvac1}), we find that supersymmetry has,
in some sense, solved the problem halfway (on a logarithmic scale).
This is encouraging, as it at least represents a step in the right
direction.  Unfortunately, it is ultimately discouraging, since
(\ref{mvac1}) was simply a guess, while (\ref{mvac2}) is actually
a reliable result in this context; supersymmetry renders the
vacuum energy finite and calculable, but the answer is still
far away from what we need.  (Subtleties in supergravity and string
theory allow us to add a negative contribution to the vacuum
energy, with which we could conceivably tune the answer to zero
or some other small number; but there is no reason for this
tuning to actually happen.)

But perhaps there is something deep about supersymmetry which
we don't understand, and our estimate $M_{\rm vac} \sim M_{\rm susy}$ is
simply incorrect.  What if instead the correct formula were
\be
  M_{\rm vac} \sim \left({M_{\rm susy} \over M_{\rm P}}\right) M_{\rm susy}\ ?
\ee
In other words, we are guessing that the supersymmetry-breaking
scale is actually the geometric mean of the vacuum scale and
the Planck scale.
Because $M_{\rm P}$ is fifteen orders of magnitude larger
than $M_{\rm susy}$, and $M_{\rm susy}$ is fifteen orders of magnitude larger
than $M_{\rm vac}$, this guess gives us the correct answer!  Unfortunately
this is simply optimistic numerology; there is no theory that
actually yields this answer (although there are speculations in
this direction \cite{banks03}).  Still, the simplicity with which we
can write down the formula allows us to dream that an improved
understanding of supersymmetry might eventually yield the 
correct result.

As an alternative to searching for some formula that gives the vacuum
energy in terms of other measurable parameters, it may be that the vacuum
energy is not a fundamental quantity, but simply our feature of our
local environment.  We don't turn to fundamental theory for an
explanation of the average temperature of the Earth's atmosphere,
nor are we surprised that this temperature is noticeably larger
than in most places in the universe; perhaps the cosmological
constant is on the same footing.  This is the idea commonly known as
the ``anthropic principle.''

To make this idea work, we need
to imagine that there are many different regions of the universe
in which the vacuum energy takes on different values; then we would
expect to find ourselves in a region which was hospitable to our
own existence.  Although most humans don't think of the
vacuum energy as playing any role in their lives, a substantially
larger value than we presently observe would either have led to
a rapid recollapse of the universe (if $\rho_{\rm vac}$ were negative)
or an inability to form galaxies (if $\rho_{\rm vac}$ were positive).
Depending on the distribution of possible values of $\rho_{\rm vac}$,
one can argue that the observed value is in excellent
agreement with what we should expect \cite{linde86,weinberg87,vilenkin95,
efstathiou95,msw98,gv00,gv03}.

The idea of environmental selection only works
under certain special circumstances, and we are far from 
understanding whether those conditions hold in our
universe.  In particular, we need to show that there can be a
huge number of different domains with slightly different values
of the vacuum energy, and that the domains can be big enough that
our entire observable universe is a single domain, and that the
possible variation of other physical quantities from domain to domain
is consistent with what we observe in ours.

Recent work in string theory has lent some support to the idea
that there are a wide variety of possible vacuum states rather
than a unique one \cite{Dasgupta:1999ss,Bousso:2000xa,Feng:2000if,
Giddings:2001yu,Kachru:2003aw,Susskind:2003kw}.  String
theorists have been investigating novel ways to compactify
extra dimensions, in which crucial roles are played by
branes and gauge fields.  By taking
different combinations of extra-dimensional geometries, brane
configurations, and gauge-field fluxes, it seems plausible that a
wide variety of states may be constructed, with different local
values of the vacuum energy and other physical parameters. 
An obstacle to understanding these purported solutions is the
role of supersymmetry, which is an important part of string theory
but needs to be broken to obtain a realistic universe.  From the
point of view of a four-dimensional observer, the compactifications
that have small values of the cosmological constant would appear to 
be exactly the states alluded to earlier, where
one begins with a supersymmetric state with a negative vacuum energy,
to which supersymmetry breaking adds just the right amount of positive
vacuum energy to give a small overall value.  The necessary
fine-tuning is accomplished simply by imagining that there are
many (more than $10^{100}$) such states, so that even very unlikely
things will sometimes occur.  We still have a long way to go 
before we understand this possibility; in particular, it is not
clear that the many states obtained have all the desired
properties \cite{Banks:2003es}.

Even if such states are allowed, it is necessary to imagine a universe
in which a large number of them actually exist in local regions
widely separated from each other.  As is well known,
inflation works to take a small region
of space and expand it to a size larger than the observable universe;
it is not much of a stretch to imagine that a multitude of different
domains may be separately inflated, each with different vacuum
energies.  Indeed, models of inflation generally tend to be
eternal, in the sense that the universe continues to inflate in
some regions even after inflation has ended in others 
\cite{Vilenkin:xq,Linde:fd}.
Thus, our observable universe may be separated by inflating regions
from other ``universes'' which have landed in different vacuum
states; this is precisely what is needed to empower the idea
of environmental selection.

Nevertheless, it seems extravagant to imagine a fantastic number
of separate regions of the universe, outside the boundary of
what we can ever possibly observe, just so that we may understand
the value of the vacuum energy in our region.  But again, this
doesn't mean it isn't true.  To decide once and for all will
be extremely difficult, and will at the least require a much
better understanding of how both string theory (or some
alternative) and inflation operate -- an understanding that
we will undoubtedly require a great deal of experimental input
to achieve.

\subsection{Dark Energy, or Worse?}

If general relativity is correct, cosmic acceleration implies there 
must be a dark energy density which diminishes relatively slowly
as the universe expands.  This can be seen directly from the Friedmann
equation (\ref{Friedmann}), which implies
\be
 {\dot a}^2 \propto a^2 \rho + {\rm constant}\ .
\ee
From this relation, it is clear that the only way to get
acceleration ($\dot a$ increasing) in an expanding universe
is if $\rho$ falls off more
slowly than $a^{-2}$; neither matter ($\rho_{\rm M} \propto a^{-3}$)
nor radiation ($\rho_{\rm R}\propto a^{-4}$) will do the trick.
Vacuum energy is, of course, strictly constant; but the data 
are consistent with smoothly-distributed sources of dark energy
that vary slowly with time.

There are good reasons to consider dynamical dark
energy as an alternative to an honest cosmological constant.
First, a dynamical energy density can be evolving slowly to zero,
allowing for a solution to the cosmological constant problem which
makes the ultimate vacuum energy vanish exactly.  Second, it poses
an interesting and challenging observational problem to study the
evolution of the dark energy, from which we might learn something
about the underlying physical mechanism.  Perhaps most intriguingly,
allowing the dark energy to evolve opens the possibility
of finding a dynamical solution to the coincidence problem, if the
dynamics are such as to trigger a recent takeover by the dark energy
(independently of, or at least for a wide range of, the 
parameters in the theory).  To date this hope has not quite been
met, but dynamical mechanisms at least allow for the possibility 
(unlike a true cosmological constant).

The simplest possibility along these lines
involves the same kind of source
typically invoked in models of inflation in the very early universe:
a scalar field $\phi$ rolling slowly in a potential, sometimes known as
``quintessence'' \cite{Wetterich:fm,Peebles:1987ek,Ratra:1987rm,
Frieman:1991tu,Caldwell:1997ii,Huey:1998se}. 
The energy density of a scalar field is a sum of kinetic, gradient,
and potential energies,
\be
 \rho_\phi = {1\over 2}{\dot\phi}^2 + {1\over 2}(\nabla\phi)^2 + 
 V(\phi)\ .
 \label{rhophi}
\ee
For a homogeneous field ($\nabla\phi \approx 0$),
the equation of motion in an expanding universe is
\be
 \ddot\phi + 3H\dot\phi + {dV \over d\phi} = 0\ .
\ee
If the slope of the potential $V$ is quite flat, we will have
solutions for which $\phi$ is nearly constant throughout space and
only evolving very gradually with time; the energy density in
such a configuration is
\be
  \rho_\phi \approx V(\phi) \approx {\rm constant}\ .
\ee
Thus, a slowly-rolling scalar field is an appropriate candidate
for dark energy.

However, introducing dynamics opens up the possibility
of introducing new problems, the form and severity
of which will depend on the specific
kind of model being considered.  Most quintessence
models feature scalar fields $\phi$ with masses of order the 
current Hubble scale,
\be
  m_\phi \sim H_0 \sim 10^{-33} {\rm ~eV}\ .
\ee
(Fields with larger masses would typically have already rolled
to the minimum of their potentials.)
In quantum field theory, light scalar fields are
unnatural; renormalization effects tend to drive scalar masses
up to the scale of new physics.  The well-known hierarchy
problem of particle physics amounts to asking why the Higgs
mass, thought to be of order $10^{11}$~eV, should be so much
smaller than the grand unification/Planck scale, 
$10^{25}$-$10^{27}$~eV.  Masses of $10^{-33}$~eV are 
correspondingly harder to understand.  On top of that, light
scalar fields give rise to long-range forces and time-dependent
coupling constants that should be observable even if couplings
to ordinary matter are suppressed by the Planck scale
\cite{Carroll:1998zi,Dvali:2001dd}; we therefore need to invoke
additional fine-tunings to explain why the quintessence field
has not already been experimentally detected.

Nevertheless, these apparent fine-tunings might be worth the
price, if we were somehow able to explain the coincidence problem.
To date, many investigations have considered scalar fields with
potentials that asymptote gradually to zero, of the form
$e^{1/\phi}$ or $1/\phi$.  These can have cosmologically interesting
properties, including ``tracking'' behavior that makes the current
energy density largely independent of the initial conditions
\cite{Zlatev:1998tr}.  They do not, however,
provide a solution to the coincidence problem, as the era in which
the scalar field begins to dominate is still set by finely-tuned
parameters in the theory.  One way to address the coincidence
problem is to take advantage of the fact that matter/radiation 
equality was a relatively recent occurrence (at least on a 
logarithmic scale); if a scalar field has dynamics which are
sensitive to the difference between matter- and radiation-dominated
universes, we might hope that its energy density becomes constant
only after matter/radiation equality.  An approach which takes this
route is $k$-essence \cite{Armendariz-Picon:2000dh}, 
which modifies the form of the kinetic energy for the scalar field.  
Instead of a conventional kinetic energy $K={1\over 2}(\dot\phi)^2$,
in $k$-essence we posit a form
\be
  K = f(\phi) g(\dot\phi^2)\ ,
\ee
where $f$ and $g$ are functions specified by the model.  
For certain choices of these functions, 
the $k$-essence field naturally tracks the evolution of
the total radiation energy density during radiation domination,
but switches to being almost constant once matter begins to
dominate.  Unfortunately,
it seems necessary to choose a finely-tuned kinetic term to get
the desired behavior \cite{Malquarti:2003hn}.

An alternative possibility is that there is nothing special about
the present era; rather, acceleration is just something that
happens from time to time.  This can be accomplished by oscillating
dark energy \cite{Dodelson:2001fq}.
In these models the potential takes the form of a decaying 
exponential (which by itself would give scaling behavior, so that
the dark energy remained proportional to the background density) with
small perturbations superimposed:
\be
  V(\phi) = e^{-\phi}[1 + \alpha\cos(\phi)]\ .
\ee
On average, the dark energy in such a model will track that of
the dominant matter/radiation component; however, there will be
gradual oscillations from a negligible density to a dominant
density and back, on a timescale set by the Hubble parameter,
leading to occasional periods of acceleration.
Unfortunately, in neither
the $k$-essence models nor the oscillating models do we have a
compelling particle-physics motivation for the chosen dynamics,
and in both cases the behavior still depends sensitively on the
precise form of parameters and interactions chosen.  Nevertheless,
these theories stand as interesting attempts to address the 
coincidence problem by dynamical means.

One of the interesting features of dynamical dark energy is that
it is experimentally testable.
\begin{figure}
\centerline{
\psfig{figure=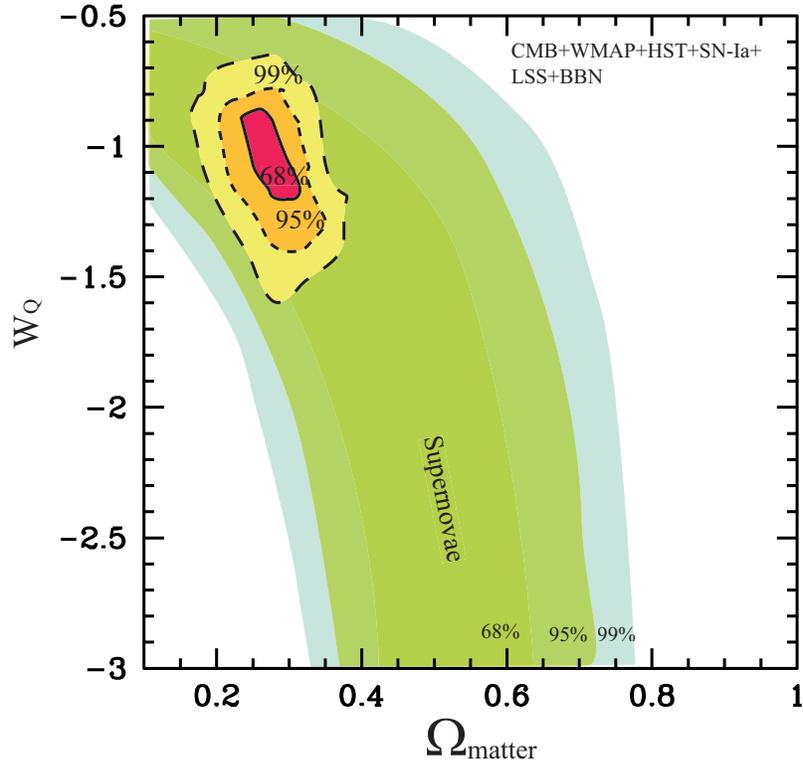,height=4in}}
\caption{Constraints on the dark-energy equation-of-state parameter,
as a function of $\Omega_{\rm M}$, assuming a flat universe.  These
limits are derived from studies of supernovae, CMB anisotropies,
measurements of the Hubble constant, large-scale structure, and
primordial nucleosynthesis.  From~\cite{Melchiorri:2002ux}.}
\label{wconstraints}
\end{figure}
In principle, different dark energy models can yield different cosmic
histories, and, in particular, a different value for the equation of
state parameter, both today and its redshift-dependence.  Since the
CMB strongly constrains the total density to be near the critical
value, it is sensible to assume a perfectly flat universe and
determine constraints on the matter density and dark energy equation
of state; see figure~(\ref{wconstraints}) for some recent limits.

As can be seen in~(\ref{wconstraints}), one possibility that is
consistent with the data is that $w<-1$. Such a possibility violates
the dominant energy condition, but possible models have been
proposed~\cite{Caldwell:1999ew}. However, such models run into serious
problems when one takes them seriously as a particle physics
theory~\cite{Carroll:2003st,Cline:2003gs}. Even if one restricts one's
attention to more conventional matter sources, making dark energy
compatible with sensible particle physics has proven tremendously
difficult.

Given the challenge of this problem, it is worthwhile considering the
possibility that cosmic acceleration is not due to some kind of stuff,
but rather arises from new gravitational physics. there are a number
of different approaches to this~\cite{Deffayet:2001pu,Freese:2002sq,
Arkani-Hamed:2002fu,Dvali:2003rk,Carroll:2003wy,Arkani-Hamed:2003uy}
and we will not review them all here. Instead we will provide an
example drawn from our own proposal~\cite{Carroll:2003wy}.

As a first attempt, consider the simplest correction to the
Einstein-Hilbert action,
\begin{equation}
\label{action} S =\frac{\mpl^2}{2}\int d^4 x\,
\sqrt{-g}\left(R-\frac{\mu^4}{R}\right) +\int d^4 x\, \sqrt{-g}\,
{\cal L}_M \ . 
\end{equation} 
Here $\mu$ is a new parameter with units
of $[{\rm mass}]$ and ${\cal L}_M$ is the Lagrangian density for matter.

The fourth-order equations arising from this action are complicated
and it is difficult to extract details about cosmological evolution
from them.  It is therefore convenient to transform from the frame
used in \ref{action}, which we call the {\em matter frame}, to the
{\em Einstein frame}, where the gravitational Lagrangian takes the
Einstein-Hilbert form and the additional degrees of freedom ($\ddot H$
and $\dot H$) are represented by a fictitious scalar field $\phi$.
The details of this can be found in~\cite{Carroll:2003wy}. Here we
just state that, performing a simultaneous redefinition of the time
coordinate, in terms of the new metric ${\tilde g}_{\mu\nu}$, our
theory is that of a scalar field $\phi(x^{\mu})$ minimally coupled to
Einstein gravity, and non-minimally coupled to matter, with potential
\begin{equation} 
\label{potential} 
V(\phi)=\mu^2 \mpl^2
\exp\left(-2\sqrt{\frac{2}{3}}\frac{\phi}{\mpl} \right)\sqrt{\exp
\left(\sqrt{\frac{2}{3}}\frac{\phi}{\mpl} \right)-1} \ .
\end{equation} 

Now let us first focus on vacuum cosmological solutions. 
The beginning of the Universe corresponds to $R
\rightarrow \infty$ and $\phi \rightarrow 0$.  The initial conditions
we must specify are the initial values of $\phi$ and $\phi'$, denoted
as $\phi_i$ and ${\phi'}_i$.  For simplicity we take $\phi_i \ll
\mpl$. There are then three qualitatively distinct outcomes, depending
on the value of ${\phi'}_i$.

{\em 1.  Eternal de Sitter.}  There is a critical value of ${\phi'}_i
\equiv {\phi'}_C$ for which $\phi$ just reaches the maximum of the
potential $V(\phi)$ and comes to rest.  In this case the Universe
asymptotically evolves to a de~Sitter solution. This
solution requires tuning and is unstable, since any perturbation will
induce the field to roll away from the maximum of its potential.

{\em 2.  Power-Law Acceleration.}  For ${\phi'}_i > {\phi'}_C$, the
field overshoots the maximum of $V(\phi )$ and the Universe evolves to
late-time power-law inflation, with observational consequences similar
to dark energy with equation-of-state parameter $w_{\rm DE}=-2/3$.

{\em 3.  Future Singularity.}  For ${\phi'}_i < {\phi'}_C$, $\phi$
does not reach the maximum of its potential and rolls back down to
$\phi =0$.  This yields a future curvature singularity.

In the more interesting case in which the Universe
contains matter, it is possible to show that the three
possible cosmic futures identified in the vacuum case remain in the
presence of matter. 

By choosing $\mu\sim 10^{-33}\,$eV, the corrections to the standard
cosmology only become important at the present epoch, making this
theory a candidate to explain the observed acceleration of the
Universe without recourse to dark energy.  Since we have no particular
reason for this choice, such a tuning appears no more attractive than
the traditional choice of the cosmological constant.

Clearly our choice of correction to the gravitational action can be
generalized.  Terms of the form $-\mu^{2(n+1)}/R^n$, with $n>1$, lead
to similar late-time self acceleration, with behavior similar to a dark energy 
component with equation of
state parameter 
\begin{equation} \label{geneos} w_{\rm eff} = -1 +
\frac{2(n+2)}{3(2n+1)(n+1)} \ . 
\end{equation} 
Clearly therefore, such modifications can
easily accommodate current observational
bounds~\cite{Melchiorri:2002ux,Spergel:2003cb} on the equation of
state parameter $-1.45< w_{\rm DE} <-0.74$ ($95\%$ confidence level).
In the asymptotic regime $n=1$ is ruled out at this level, while
$n\geq 2$ is allowed; even $n=1$ is permitted if we are near
the top of the potential.

Finally, any modification of the Einstein-Hilbert action must, of
course, be consistent with the classic solar system tests of gravity
theory, as well as numerous other astrophysical dynamical tests. 
We have chosen the coupling constant $\mu$ to be very small, but
we have also introduced a new light degree of freedom.  
Chiba \cite{Chiba:2003ir} has pointed out
that the model with $n=1$ is equivalent to Brans-Dicke
theory with $\omega=0$ in the approximation where the potential
was neglected, and would therefore be inconsistent with experiment.
It is not yet clear whether including the potential, or considering
extensions of the original model, could alter this conclusion.

\section{Early Times in the Standard Cosmology}

In the first lecture we described the kinematics and dynamics
of homogeneous and isotropic cosmologies in general relativity,
while in the second we discussed the situation in our current
universe.  In this lecture we wind the clock back, using what
we know of the laws of physics and the universe today to infer
conditions in the early universe.  Early times were characterized
by very high temperatures and densities, with many particle
species kept in (approximate) thermal equilibrium by rapid
interactions.  We will therefore have to move beyond a simple
description of non-interacting ``matter'' and ``radiation,'' and
discuss how thermodynamics works in an expanding universe.

\subsection{Describing Matter}
In the first lecture we discussed how to describe matter as a
perfect fluid, described by an energy-momentum tensor
\begin{equation}
T_{\mu\nu} = (\rho + p)U_\mu U_\nu + p g_{\mu\nu}\ ,
\end{equation}
where $U^{\mu}$ is the fluid four-velocity, $\rho$ is the energy
density in the rest frame of the fluid and $p$ is the pressure in that
same frame.  The energy-momentum tensor is covariantly conserved,
\begin{equation}
\label{emtensorconservation}
\nabla_{\mu}T^{\mu\nu}=0 \ .
\end{equation}

In a more complete description, a fluid will be characterized
by quantities in addition to the energy density and pressure.
Many fluids have a conserved quantity associated with them
and so we will also introduce a {\it number flux
density} $N^{\mu}$, which is also conserved
\begin{equation}
\nabla_{\mu}N^{\mu}=0 \ .
\end{equation}
For non-tachyonic matter $N^{\mu}$ is a timelike 4-vector and
therefore we may decompose it as
\begin{equation}
N^{\mu}=nU^{\mu} \ .
\end{equation}
We can also introduce
an {\it entropy flux density} $S^{\mu}$.  This quantity is not
conserved, but rather obeys a covariant version of the second law of
thermodynamics
\begin{equation}
\label{2ndlaw}
\nabla_{\mu}S^{\mu} \geq 0 \ .
\end{equation}
Not all phenomena are successfully described in terms of such a
local entropy vector (e.g., black holes); fortunately, it suffices
for a wide variety of fluids relevant to cosmology.

The conservation law for the energy-momentum tensor yields, most
importantly, equation~(\ref{energyconservation}), which can be thought
of as the first law of thermodynamics
\begin{equation}
\label{firstlaw}
dU=TdS-pdV \ ,
\end{equation}
with $dS=0$.

It is useful to resolve $S^{\mu}$ into components parallel and
perpendicular to the fluid 4-velocity
\begin{equation}
\label{resolveentropy}
S^{\mu}=sU^{\mu} +s^{\mu} \ ,
\end{equation}
where $s_{\mu}U^{\mu}=0$.  
The scalar $s$ is the rest-frame entropy density which, up to an
additive constant (that we can consistently set to zero),
can be written as
\be
  s = {\rho + p \over T}\ .
  \label{entropydensity}
\ee
In addition to all these quantities, we must specify an equation of
state, and we typically do this in such a way as to treat $n$ and $s$
as independent variables.

\subsection{Particles in Equilibrium}

The various particles inhabiting the early universe can be usefully
characterized according to three criteria:  in equilibrium vs.\ 
out of equilibrium (decoupled), bosonic vs.\ fermionic, and 
relativistic (velocities near $c$) vs.\ non-relativistic.  In this
section we consider species which are in equilibrium with the
surrounding thermal bath.

Let us begin by discussing the conditions under which a particle
species will be in equilibrium with the surrounding thermal plasma.
A given species remains in thermal equilibrium as long as its
interaction rate is larger than the expansion rate of the
universe. Roughly speaking, equilibrium requires it to be possible for
the products of a given reaction have the opportunity to recombine in
the reverse reaction and if the expansion of the universe is rapid
enough this won't happen. A particle species for which the interaction
rates have fallen below the expansion rate of the universe is said to
have {\it frozen out} or {\it decoupled}.  If the interaction rate of
some particle with the background plasma is $\Gamma$, it will
be decoupled whenever
\be
  \Gamma \ll H\ ,
\ee
where the Hubble constant $H$ sets the cosmological timescale.

As a good rule of thumb, the expansion rate in the early universe
is ``slow,'' and particles tend to be in thermal equilibrium (unless
they are very weakly coupled).  This
can be seen from the Friedmann equation when the energy density
is dominated by a plasma with $\rho \sim T^4$; we then have
\be
  H \sim \left({T\over \mpl}\right)T\ .
\ee
Thus, the Hubble parameter is suppressed with respect to the
temperature by a factor of $T/\mpl$.  At extremely early times
(near the Planck era, for example), the universe may be expanding
so quickly that no species are in equilibrium; as the expansion
rate slows, equilibrium becomes possible.  However, the interaction
rate $\Gamma$ for a particle with cross-section $\sigma$
is typically of the form
\be
  \Gamma = n\langle \sigma v\rangle\ ,
\ee
where $n$ is the number density and $v$ a typical particle velocity.
Since $n \propto a^{-3}$, the density of particles will eventually
dip so low that equilibrium can once again no longer be maintained.  In our
current universe, no species are in equilibrium with the background
plasma (represented by the CMB photons).

Now let us focus on particles in equilibrium.
For a gas of weakly-interacting particles, we can describe the
state in terms of a {\it distribution function} $f({\bf p})$,
where the three-momentum ${\bf p}$ satisfies
\be
  E^2({\bf p}) = m^2 + |{\bf p}|^2\ .
\ee
The distribution function characterizes the density of particles
in a given momentum bin.  (In general it will also be a function
of the spatial position ${\bf x}$, but we suppress that here.)
The number density, energy density, and pressure of some species
labeled $i$ are given by
\bea
  n_i &=& {g_i \over (2\pi)^3} \int f_i({\bf p}) d^3p \cr
  \rho_i &=& {g_i \over (2\pi)^3} \int E({\bf p})f_i({\bf p}) d^3p \cr
  p_i &=& {g_i \over (2\pi)^3} \int 
  {|{\bf p}|^2 \over 3E({\bf p})}f_i({\bf p}) d^3p \ ,
\eea
where $g_i$ is the number of spin states of the particles.  For 
massless photons we have $g_\gamma=2$, while for a massive vector
boson such as the $Z$ we have $g_Z = 3$.  In
the usual accounting, particles and antiparticles are treated as
separate species; thus, for spin-1/2 electrons and positrons we have
$g_{e^-} = g_{e^+} = 2$.  In thermal equilibrium at a temperature
$T$ the particles will be in either
Fermi-Dirac or Bose-Einstein distributions,
\be
  f({\bf p}) = {1 \over e^{E({\bf p})/T} \pm 1}\ ,
\ee
where the plus sign is for fermions and the minus sign for bosons.

We can do the integrals over the distribution functions in
two opposite limits:  particles which are highly relativistic
$(T\gg m)$ or highly non-relativistic $(T\ll m)$.
The results are shown in table~2, in which $\zeta$
is the Riemann zeta function, and $\zeta(3)\approx 1.202$.
\begin{table}
\begin{center}
\begin{tabular}{r|c|c|c}
& {Relativistic} & {Relativistic} & {Non-relativistic} \\
& {Bosons} & {Fermions} & ({Either}) \\ 
\hline &&& \\ 
$n_i$ & ${\zeta(3)\over \pi^2} g_i T^3$ & 
$\left({3\over 4}\right){\zeta(3)\over \pi^2} g_i T^3$ & 
$g_i \left({m_i T \over 2\pi}\right)^{3/2}e^{-m_i/T}$ \\
&&& \\
$\rho_i$ & ${\pi^2 \over 30} g_i T^4$ & 
$\left({7\over 8}\right){\pi^2 \over 30} g_i T^4$ & 
$m_in_i$ \\
&&& \\
$p_i$ & ${1\over 3}\rho_i$ & ${1\over 3}\rho_i$ & 
$n_iT \ll \rho_i$ \\
\end{tabular}
\end{center}
\caption{Number density, energy density, and pressure, for
 species in thermal equilibrium.}
\end{table}

From this table we can extract several pieces of relevant
information.  Relativistic particles, whether bosons or fermions,
remain in approximately equal abundances in equilibrium.  Once
they become non-relativistic, however, their abundance plummets,
and becomes exponentially suppressed with respect to the relativistic
species.  This is simply because it becomes progressively harder
for massive particle-antiparticle pairs to be produced in a plasma
with $T \ll m$.

It is interesting to note that, although matter is much more dominant
than radiation in the universe today, since their energy densities
scale differently the early universe was radiation-dominated.
We can write the ratio of the density parameters in matter and radiation
as
\be
  {\Omega_{\rm M}\over \Omega_{\rm R}} =
  {\Omega_{\rm M0}\over \Omega_{\rm R0}}\left({a\over a_0}\right)
  = {\Omega_{\rm M0}\over \Omega_{\rm R0}} (1+z)^{-1}\ .
\ee
The redshift of matter-radiation equality is thus
\begin{equation}
\label{zeq}
1+z_{\rm eq} = {\Omega_{\rm M0}\over \Omega_{\rm R0}}
\approx 3\times 10^3 \ .
\end{equation}
This expression assumes that the particles that are non-relativistic
today were also non-relativistic at $z_{\rm eq}$; this should be a
safe assumption, with the possible exception of massive neutrinos,
which make a minority contribution to the total density.

As we mentioned in our discussion of the CMB in the previous
lecture, even decoupled photons maintain a thermal distribution;
this is not because they are in equilibrium, but simply because
the distribution function redshifts into a similar distribution
with a lower temperature proportional to $1/a$.  
We can therefore speak of the ``effective
temperature'' of a relativistic species that freezes out at
a temperature $T_f$ and scale factor $a_f$:
\be
  T_i^{\rm rel}(a) = T_f \left({a_f \over a}\right)\ .
\ee
For example, neutrinos decouple at a temperature around
1~MeV; shortly thereafter, electrons and positrons annihilate
into photons, dumping energy (and entropy) into the plasma but
leaving the neutrinos unaffected.  Consequently, we expect a
neutrino background in the current universe with a temperature
of approximately 2K, while the photon temperature is 3K.

A similar effect occurs for particles which are non-relativistic
at decoupling, with one important difference.  For non-relativistic
particles the temperature is proportional to the kinetic energy
${1\over 2}mv^2$, which redshifts as $1/a^2$.  We therefore have
\be
  T_i^{\rm non-rel}(a) = T_f \left({a_f \over a}\right)^2\ .
\ee
In either case we are imagining that the species freezes out while
relativistic/non-relativistic and stays that way afterward; if 
it freezes out while relativistic and subsequently becomes 
non-relativistic, the distribution function will be distorted
away from a thermal spectrum.

The notion of an effective temperature allows us to define a
corresponding notion of an effective number of relativistic 
degrees of freedom, which in turn permits a compact expression for the
total relativistic energy density.  The effective number of 
relativistic degrees of freedom (as far as energy is concerned) 
can be defined as
\be
  g_* = \sum_{\rm bosons} g_i\left(T_i \over T\right)^4
  + {7\over 8}\sum_{\rm fermions} g_i\left(T_i \over T\right)^4\ .
\ee
(The temperature $T$ is the actual temperature of the background
plasma, assumed to be in equilibrium.)
Then the total energy density in all relativistic species 
comes from adding the contributions
of each species, to obtain the simple formula
\be
  \rho = {\pi^2 \over 30} g_* T^4\ .
\ee
We can do the same thing for the entropy density.  From 
(\ref{entropydensity}), the entropy density in relativistic particles
goes as $T^3$ rather than $T^4$, so we define the effective number
of relativistic degrees of freedom for entropy as
\be
  g_{*S} = \sum_{\rm bosons} g_i\left(T_i \over T\right)^3
  + {7\over 8}\sum_{\rm fermions} g_i\left(T_i \over T\right)^3\ .
\ee
The entropy density in relativistic species is then
\be
  s = {2\pi \over 45}g_{*S} T^3 \ .
  \label{entropydensity2}
\ee
Numerically, $g_*$ and $g_*{S}$ will typically be very close to
each other.  In the Standard Model, we have
\be
  g_* \approx g_{*S} \sim \cases{ 100 & $T > 300$~MeV \cr
  10 & $300 {\rm ~MeV} > T > 1$~MeV \cr 3 & $T < 1$~MeV\ . }
  \label{8.161}
\ee
The events that change the effective
number of relativistic degrees of freedom are the QCD phase
transition at 300~MeV, and the annihilation of electron/positron
pairs at 1~MeV. 

Because of the release of energy into the background plasma when
species annihilate, it is only an approximation to say that the
temperature goes as $T\propto 1/a$.  A better approximation is to
say that the comoving entropy density is conserved,
\be
  s \propto a^{-3}\ .
\ee
This will hold under all forms of adiabatic evolution; entropy
will only be produced at a process like a first-order phase transition 
or an out-of-equilibrium decay.  (In fact, we expect that the entropy
production from such processes is very small compared to the total
entropy, and adiabatic evolution is an excellent approximation for
almost the entire early universe.  One exception is inflation,
discussed in the next lecture.)  Combining entropy conservation 
with the expression
(\ref{entropydensity2}) for the entropy density in relativistic
species, we obtain a better expression for the evolution of the
temperature,
\be
  T \propto g_{*S}^{-1/3}a^{-1}\ .
\ee
The temperature will consistently decrease under adiabatic evolution
in an expanding universe, but it decreases more slowly when the
effective number of relativistic degrees of freedom is diminished.

\subsection{Thermal Relics}

As we have mentioned, particles typically do not stay in 
equilibrium forever; eventually the density becomes so low that
interactions become infrequent, and the particles freeze out.  Since
essentially all of the particles in our current universe fall into this
category, it is important to study the relic abundance of decoupled
species.  (Of course it is also possible to obtain a significant
relic abundance for particles which were never in thermal 
equilibrium; examples might include baryons produced by GUT
baryogenesis, or axions produced by vacuum misalignment.)  In this
section we will typically neglect factors of order unity.

We have seen that relativistic, or {\it hot}, particles have a number
density that is proportional to $T^3$ in equilibrium. Thus, a
species $X$ that freezes out while still relativistic will have a
number density at freeze-out $T_f$ given by
\begin{equation}
n_X(T_f)\sim T_f^3 \ .
\end{equation}
Since this is comparable to the number density of photons at that
time, and after freeze-out both photons and our species $X$ just have
their number densities dilute by a factor $a(t)^{-3}$ as the universe
expands, it is simple to see that the abundance of $X$ particles today
should be comparable to the abundance of CMB photons,
\be
  n_{X0} \sim n_{\gamma 0} \sim 10^2~{\rm cm}^{-3}\ .
\ee
We express this number as $10^2$ rather than $411$ since the
roughness of our estimate does not warrant
such misleading precision.  The leading correction to this value
is typically due to the production of additional photons
subsequent to the decoupling of $X$; in the Standard Model,
the number density of photons increases by a factor of
approximately $100$ between the electroweak phase transition
and today, and a species which decouples during this period
will be diluted by a factor of between $1$ and $100$ depending
on precisely when it freezes out.  So, for example, neutrinos 
which are light
($m_\nu < $~MeV) have a number density today of $n_\nu = 
115$~cm$^{-3}$ per species, and a corresponding contribution to the 
density parameter (if they are nevertheless heavy enough to be
nonrelativistic today) of
\be
  \Omega_{0, \nu} = \left({{m_\nu}\over{92~{\rm eV}}}\right) h^{-2}.
\ee
(In this final expression we have secretly taken account of the
missing numerical factors, so this is a reliable answer.)
Thus, a neutrino with $m_\nu\sim 10^{-2}$~eV
would contribute $\Omega_\nu\sim 2\times 10^{-4}$.
This is large enough to be interesting without being large enough
to make neutrinos be the dark matter.  That's good news, since
the large velocities of neutrinos make them free-stream out of
overdense regions, diminishing primordial perturbations and
leaving us with a universe which has much less structure on
small scales than we actually observe.

Now consider instead a species $X$ which is nonrelativistic or {\it
cold} at the time of decoupling.  It is much harder to accurately
calculate the relic abundance of a cold relic than a hot one,
simply because the equilibrium abundance of a nonrelativistic
species is changing rapidly with respect to the background plasma,
and we have to be quite precise following the freeze-out
process to obtain a reliable answer.  The accurate calculation
typically involves numerical integration of the Boltzmann equation
for a network of interacting particle species; here, we cut to the
chase and simply provide a reasonable approximate expression.  If
$\sigma_0$ is the annihilation cross-section of the species $X$
at a temperature $T=m_X$, the final number density in terms of
the photon density works out to be 
\begin{equation}
n_X(T<T_f)  \sim \frac{1}{\sigma_0 m_X M_P} n_{\gamma}\ .
\end{equation}
Since the particles are nonrelativistic when they decouple, they will
certainly be nonrelativistic today, and their energy density is 
\be
  \rho_X = m_X n_X\ .
\ee
We can plug in numbers for the Hubble parameter and photon density
to obtain the density parameter,
\be
  \Omega_X = {\rho_X \over \rho_{\rm cr}} \sim 
  {n_\gamma \over \sigma_0 \mpl^3 H_0^2}\ .
\ee
Numerically, when $\hbar=c=1$ we have 1~GeV$\sim 2\times 10^{-14}$~cm,
so the photon density today is $n_\gamma \sim 100$~cm$^{-3} \sim
10^{-39}$~GeV$^{-3}$.  The Hubble constant is $H_0\sim 10^{-42}$~GeV,
and the Planck mass is $\mpl \sim 10^{18}$~GeV, so we obtain
\be
  \Omega_X \sim 
  {1 \over \sigma_0(10^{9}~{\rm GeV}^2)}\ .
\ee
It is interesting to note that this final expression is independent
of the mass $m_X$ of our relic, and only depends on the annihilation
cross-section; that's because more massive particles will have a
lower relic abundance.  Of course, this depends on how we choose to
characterize our theory; we may use variables in which $\sigma_0$ is
a function of $m_X$, in which case it is reasonable to say that the
density parameter does depend on the mass. 

The designation {\it cold} may ring a bell with many of you, for you
will have heard it used in a cosmological context applied to {\it cold
dark matter} (CDM). Let us see briefly why this is. One candidate for
CDM is a Weakly Interacting Massive Particle (WIMP). The annihilation
cross-section of these particles, since they are weakly interacting,
should be $\sigma_0\sim \alpha^2_W G_F$, where $\alpha_W$ is the
weak coupling constant and $G_F$ is the 
the Fermi constant.  Using $G_F\sim (300 {\rm GeV})^{-2}$  
and $\alpha_W\sim 10^{-2}$, we get
\be
  \sigma_0 \sim \alpha^2_W G_F \sim 10^{-9}~{\rm GeV}^{-2}\ .
\ee
Thus, the density parameter in such particles 
would be
\begin{equation}
  \Omega_X \sim 1\ .
\end{equation}
In other words, a stable particle with a weak
interaction cross section naturally produces a relic density of order
the critical density today, and so provides a perfect candidate for
cold dark matter.  A paradigmatic example is provided by the lightest supersymmetric
partner (LSP), if it is stable and supersymmetry is broken at the
weak scale.  Such a possibility is of great interest to both particle physicists and cosmologists, since it
may be possible to produce and detect such particles in colliders and 
to directly detect a WIMP background in cryogenic detectors
in underground laboratories; this will be a major experimental effort
over the next few years \cite{Akerib:2002ez}.

\subsection{Vacuum displacement}
\label{subsec:vacdis}

Another important possibility is the existence of relics
which were never in thermal equilibrium.  An example of these will be
discussed later in this lecture:  the production of topological defects
at phase transitions.  Let's discuss another kind of 
non-thermal relic, which derives from what we might call
``vacuum displacement.''  Consider the action for a real scalar
field in curved spacetime:
\be
  S = \int d^4x\, \sqrt{-g}\left[-{1\over 2}g^{\mu\nu}
  \partial_\mu\phi \partial_\nu\phi - V(\phi)\right]\ .
\ee
If we assume that $\phi$ is spatially homogeneous ($\partial_i\phi
= 0$), its equation of motion in
the Robertson-Walker metric (\ref{frwmetric}) will be
\be
  \ddot\phi + 3H\dot\phi + V'(\phi) = 0\ ,\label{phifrw}
\ee
where an overdot indicates a partial derivative with respect
to time, and a prime indicates a derivative with respect to $\phi$.
For a free massive scalar field, $V(\phi) = {1\over 2} m_\phi^2
\phi^2$, and (\ref{phifrw}) describes a harmonic oscillator with
a time-dependent damping term.  For $H > m_\phi$ the
field will be overdamped, and stay essentially constant at 
whatever point in the potential it finds itself.  So let us
imagine that at some time in the very early universe (when $H$
was large) we had such an overdamped homogeneous scalar field,
stuck at a value $\phi = \phi_*$; the total energy density in the field 
is simply the potential energy ${1\over 2}m_\phi^2\phi_*^2$.  
The Hubble parameter $H$ will decrease to approximately $m_\phi$ 
when the temperature reaches $T_*=\sqrt{m_\phi \mpl}$, 
after which the field
will be able to evolve and will begin to oscillate in its
potential.  The vacuum energy is converted to a combination of
vacuum and kinetic energy which will redshift like matter,
as $\rho_\phi\propto a^{-3}$; in a particle interpretation, the
field is a Bose condensate of zero-momentum particles.  We
will therefore have
\be
  \rho_\phi(a) \sim {1\over 2}m_\phi^2 \phi_*^2\left(
  {{a_*}\over a}\right)^3\ ,
\ee
which leads to a density parameter today
\be
  \Omega_{0, \phi} \sim \left({{\phi_*^4 m_\phi}\over{10^{-19}
  ~{\rm GeV}^5}}\right)^{1/2}\ .
  \label{omegadisp}
\ee

A classic example of a non-thermal relic produced by vacuum 
displacement is the QCD axion, which has a typical primordial value
$\langle\phi\rangle\sim f_{\rm PQ}$ and a mass $m_\phi \sim
\Lambda_{\rm QCD}^2/f_{\rm PQ}$, where $f_{\rm PQ}$ is the
Peccei-Quinn symmetry-breaking scale and $\Lambda_{\rm QCD}\sim
0.3$~GeV is the QCD scale \cite{kt}.  In this case, plugging
in numbers reveals
\be
  \Omega_{0, \phi} \sim \left({{f_{\rm PQ}}\over{10^{13}~{\rm GeV}}}
  \right)^{3/2}\ .
  \label{vacdisomega}
\ee
The Peccei-Quinn scale is essentially a free parameter from a
theoretical point of view, but experiments and astrophysical
constraints have ruled out most values except for a small window
around $f_{\rm PQ}\sim 10^{12}~{\rm GeV}$.  The axion therefore
remains a viable dark matter candidate 
\cite{Kolb:1998kj,Ellis:1998gt}.  Note that, even though 
dark matter axions are very light
($\Lambda_{\rm QCD}^2/f_{\rm PQ}\sim 10^{-4}$~eV), they are
extremely non-relativistic, which can be traced to the non-thermal
nature of their production process.  (Another important way to
produce axions is through the decay of axion cosmic strings
\cite{kt,vilenkinshellard}.)

\subsection{Primordial Nucleosynthesis}
Given our time constraints, even some of the more important concepts
in cosmology cannot be dealt with in significant detail. We have
chosen just to give a cursory treatment to primordial nucleosynthesis,
although its importance as a crucial piece of evidence in favor of the
big bang model and its usefulness in bounding any new physics of
cosmological relevance cannot be overstated.

Observations of primordial nebulae reveal abundances of the light
elements unexplained by stellar nucleosynthesis.  Although it does a
great disservice to the analytic and numerical work required, not to
mention the difficulties of measuring the abundances, we will just
state that the study of nuclear processes in the background of an
expanding cooling universe yields a remarkable concordance between
theory and experiment.

At temperatures below 1~MeV, the weak interactions are
frozen out and
neutrons and protons cease to interconvert.  The equilibrium
abundance of neutrons at this temperature is about $1/6$ the
abundance of protons (due to the slightly larger neutron mass).
The neutrons have a finite lifetime ($\tau_n = 890$~sec) that
is somewhat larger than the age of the universe at this epoch,
$t(1~{\rm MeV})\approx 1$~sec, but they begin
to gradually decay into protons and leptons.  Soon thereafter,
however, we reach a temperature somewhat below $100$~keV, and
Big-Bang nucleosynthesis (BBN) begins.  (The nuclear binding energy
per nucleon is typically of order 1~MeV, so you might expect that
nucleosynthesis would occur earlier; however, the large number of
photons per nucleon prevents nucleosynthesis from taking place
until the temperature drops below $100$~keV.)  At that point
the neutron/proton ratio is approximately $1/7$.  Of all the light
nuclei, it is energetically favorable for the nucleons to reside
in $^4$He, and indeed that is what most of the free neutrons are
converted into; for every two neutrons and fourteen protons, we
end up with one helium nucleus and twelve protons.  Thus, about
$25\%$ of the baryons by mass are converted to helium.  In addition,
there are trace amounts of deuterium (approximately $10^{-5}$
deuterons per proton), $^3$He (also $\sim 10^{-5}$), and 
$^7$Li ($\sim 10^{-10}$).

Of course these numbers are predictions, which are borne out by
observations of the primordial abundances of light elements.
(Heavier elements are not synthesized in the Big Bang, but require
supernova explosions in the later universe.)  We have glossed over
numerous crucial details, especially those which explain how the
different abundances depend on the cosmological parameters.  For
example, imagine that we deviate from the Standard Model by 
introducing more than three light neutrino species.  This would
increase the radiation energy density at a fixed temperature,
which in turn decreases the timescales
associated with a given temperature (since $t\sim H^{-1}
\propto \rho_{\rm R}^{-1/2}$).  Nucleosynthesis would therefore
happen somewhat earlier, resulting in a higher abundance of
neutrons, and hence in a larger abundance of $^4$He.  Observations
of the primordial helium abundance, which are consistent with the
Standard Model prediction, provided the first evidence that the
number of light neutrinos is actually three.  

\begin{figure}[t]
\centerline{
\psfig{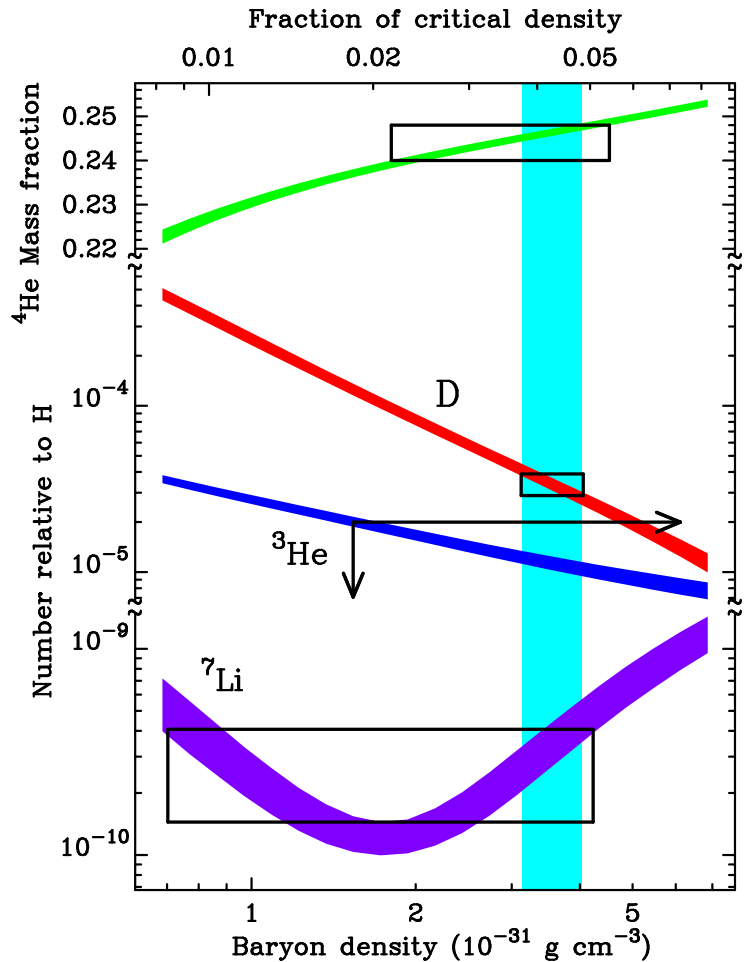}}
\caption{Abundances of light elements produced by BBN, as a function
of the baryon density.  The vertical strip indicates the concordance
region favored by observations of primordial abundances. From
\cite{Burles:1999ac}.}
\label{bbnfig}
\end{figure}

The most amazing fact about nucleosynthesis is that, given that the
universe is radiation dominated during the relevant epoch (and the
physics of general relativity and the Standard Model), the
relative abundances of the light elements depend essentially on just
one parameter, the {\it baryon to entropy ratio} \be
\label{eta}
\eta\equiv \frac{n_B}{s} = {n - n_{\bar b} \over s}\ ,
\ee
where $n_B=n_b-n_{{\bar b}}$ is the difference between the number of baryons 
and antibaryons per unit volume. 
The range of $\eta$ consistent with the deuterium and $^3$He primordial 
abundances is 
\begin{equation}
2.6\times 10^{-10} < \eta < 6.2\times 10^{-10} \ .
\end{equation}
Very recently this number has been independently determined to be 
\begin{equation}
\eta =  6.1\times 10^{-10}\ ^{+0.3\times 10^{-10}}_{-0.2\times 10^{-10}}
\end{equation} 
from precise measurements of the relative heights of the first two
microwave background (CMB) acoustic peaks by the WMAP satellite. This
is illustrated in figure~(\ref{bbnfig}) and we will have a lot more to
say about this quantity later when we discuss baryogenesis.

\subsection{Finite Temperature Phase Transitions}
We have hinted at the need to go beyond perfect fluid sources
for the Einstein equations if we are to unravel some of the mysteries
left by the standard cosmology. Given our modern understanding of
particle physics, it is natural to consider using field theory to
model matter at early times in the universe. The effect of cosmic
expansion and the associated thermodynamics yield some fascinating
phenomena when combined with such a field theory approach. One
significant example is provided by finite temperature phase
transitions.

Rather than performing a detailed calculation in finite-temperature
field theory, we will illustrate this with a rough argument. Consider
a theory of a single real scalar field $\phi$ at zero temperature,
interacting with a second real scalar field $\chi$. The Lagrangian
density is
\begin{equation}
{\cal L}=-\frac{1}{2}\partial_{\mu}\phi \partial^{\mu}\phi - \frac{1}{2}\partial_{\mu}\chi \partial^{\mu}\chi - V(\phi,T=0) -\frac{g}{2}\phi^2\chi^2\ ,
\end{equation}
with potential 
\begin{equation}
V(\phi,T=0)=-\frac{\mu^2}{2}\phi^2 +\frac{\lambda}{4}\phi^4 \ ,
\end{equation}
where $\mu$ is a parameter with dimensions of mass and $\lambda$ and
$g$ are dimensionless coupling constants.

Now, consider what the effective theory for $\phi$ looks like if we
assume that $\chi$ is in thermal equilibrium. In this case, we may
replace $\chi$ by the temperature $T$ to obtain a Lagrangian for
$\phi$ only, with {\it finite temperature effective potential} given
by
\begin{equation}
\label{fteffpot}
V(\phi,T)=\frac{1}{2}\left(gT^2-\mu^2\right)\phi^2
+\frac{\lambda}{4}\phi^4 \ .
\end{equation}

This simple example demonstrates a very significant result. At zero
temperature, all your particle physics intuition tells you correctly
that the theory is spontaneously broken, in this case the global 
${\cal Z}_2$
symmetry of the Lagrangian is broken by the ground state value
$\langle\phi\rangle \neq 0$ (we shall see more of this soon). However,
at temperatures above the critical temperature $T_c$ given by
\begin{equation}
\label{criticalT}
T_c=\frac{\mu}{\sqrt{g}} \ ,
\end{equation}
the full ${\cal Z}_2$ symmetry of the Lagrangian is respected by the finite
temperature ground state, which is now at $\langle\phi\rangle=0$.
This behavior is know as finite temperature symmetry restoration, and
its inverse, occurring as the universe cools, is called finite
temperature spontaneous symmetry breaking.

Zero temperature unification and symmetry breaking are fundamental features 
of modern particle physics models. For particle 
physicists this amounts to the 
statement that the chosen vacuum state of a gauge field theory is one which 
does not respect the underlying symmetry group of the Lagrangian. 

The melding of ideas from particle physics and cosmology described
above leads us to speculate that matter in the early universe was
described by a unified gauge field theory based on a simple continuous
Lie group, $G$. As a result of the extreme temperatures of the early
universe, the vacuum state of the theory respected the full symmetry
of the Lagrangian.  As the universe cooled, we hypothesize that the
gauge theory underwent a series of {\it spontaneous symmetry
breakings} (SSB) until matter was finally described by the unbroken
gauge groups of QCD and QED.  Schematically the symmetry breaking can
be represented by
\be
G\rightarrow H \rightarrow \cdots \rightarrow SU(3)_c \times SU(2)_L
\times U(1)_Y \rightarrow SU(3)_c \times U(1)_{em}\ .
\ee
The group $G$ is known as the grand unified gauge group and the initial
breaking $G\rightarrow H$ and is expected to
take place at $10^{16}$GeV, as we discuss below.

Let us briefly 
set up the mathematical description of SSB which will be useful later when we discuss topological properties of the theory.

Consider a gauge field theory described by a continuous group $G$.
Denote the vacuum state of the theory by $|0\rangle$. Then, given 
$g(x)\in G$, the state $g(x)|0\rangle$ is also a vacuum. Suppose all
vacuum states are of the form $g|0\rangle$ but that the state
$|0\rangle$ is invariant only under a subgroup $H\subset G$. Then the 
symmetry of the theory is said to have been spontaneously broken from
$G$ to $H$. Let us define two vacuum states $|0\rangle_A$ and
$|0\rangle_B$ to represent the same state under the broken group if
$g|0\rangle_A=|0\rangle_B$ for some $g\in G$. We write 
$|0\rangle_A \sim |0\rangle_B$ and the distinguishable vacua of the
theory are equivalence classes under $\sim$ and are the cosets of $H$
in $G$. Then the {\it vacuum manifold}, the space of all accessible 
vacua of the theory, is the coset space
\be
{\cal M}= G/H\ .
\ee
Given that we believe the universe evolved through a sequence of 
such symmetry 
breakings, there are many questions we can ask about the cosmological 
implications of the scheme and many ways in which we can constrain and 
utilize the possible breaking patterns. 

\subsection{Topological Defects}

In a quantum field theory, small oscillations around the vacuum
appear as particles.  If the space of possible vacuum states is
topologically nontrivial, however, there arises the possibility of
another kind of solitonic object:  a {\it topological defect}.
Consider a field theory described by a continuous symmetry group $G$ which 
is spontaneously broken to a subgroup $H\subset G$. Recall that the space of
all accessible vacua of the theory, the {\it vacuum manifold}, is defined to
be the space of cosets of $H$ in $G$; ${\cal M} \equiv G/H$. 
Whether the theory admits topological defects depends on whether
the vacuum manifold has nontrivial homotopy groups.  A homotopy
group consists of equivalence classes of maps of spheres (with fixed
base point) into the manifold, where two maps are equivalent if they
can be smoothly deformed into each other.  The homotopy groups
defined in terms of $n$-spheres are denoted $\pi_n$.

In general, a field theory with vacuum manifold ${\cal M}$ 
possesses a topological defect of some type if
\be 
\pi_i({\cal M}) \neq 1\ ,
\ee
for some $i=0,1,\ldots$. In particular, we can have a set of 
defects, as listed in table~3.
\begin{table}[htb]
\begin{center}
\begin{tabular}{c|c}
{Homotopy Constraint} & {Topological Defect} \\ \hline
$\pi_0({\cal M}) \neq 1$ & Domain Wall \\
$\pi_1({\cal M}) \neq 1$ & Cosmic String \\
$\pi_2({\cal M}) \neq 1$ & Monopole \\
$\pi_3({\cal M}) \neq 1$ & Texture\\ \hline
\end{tabular}
\end{center}
\caption{Topological defects, as described by homotopy groups of
the vacuum manifolds of a theory with spontaneously broken symmetry.}
\end{table}
In order to get an intuitive picture of the meaning of these topological
criteria let us consider the first two.

If $\pi_0({\cal M})\neq 1$, the manifold ${\cal M}$ is
disconnected. (A zero-sphere is the set of two points a fixed distance
from the origin in ${\bf R}^1$.  The set of topologically equivalent
maps with fixed base point from such such a sphere into a manifold is
simply the set of disconnected pieces into which the manifold falls.)
Let's assume for simplicity that the vacuum manifold consists
of just two disconnected components ${\cal M}_1$ and ${\cal M}_2$ and restrict
ourselves to one spatial dimension. Then, if we apply boundary conditions that
the vacuum at $-\infty$ lies in ${\cal M}_1$ and that at $+\infty$ lies in
${\cal M}_2$, by continuity there must be a point somewhere where the 
order parameter does not lie in the vacuum manifold as the field interpolates
between the two vacua. The region in which the field is out of the vacuum is
known as a {\it domain wall}.

Similarly, suppose $\pi_1({\cal M})\neq 1$. This implies that the
vacuum manifold is not simply-connected: there are
non-contractible loops in the manifold. The order parameter for such a
situation may be considered complex and if, when traveling around a
closed curve in space, the phase of the order parameter changes by a
non-zero multiple of $2\pi$, by continuity there is a point
within the curve where the field is out of the vacuum
manifold. Continuity in the direction perpendicular to the plane of
the curve implies that there exists a line of such points. This is an
example of a cosmic string or line defect. Since we shall mostly
concentrate on these defects let us now give a detailed analysis of
the simplest example, the {\it Nielsen-Olesen vortex} in the {\it
Abelian Higgs model}.

Consider a complex scalar field theory based on the Abelian gauge group $U(1)$.
The Lagrangian density for this model is
\be
{\cal L}=-\frac{1}{2}(D_{\mu}\vp)^*D^{\mu}\vp - \frac{1}{4}F_{\mu\nu}
F^{\mu\nu} -V(\vp)\ ,
\ee
where $\vp(x)$ is the complex scalar field, the covariant derivative,
$D_{\mu}$, and field strength tensor $F_{\mu\nu}$ are defined
in terms of the Abelian gauge field $A_{\mu}$ as 
\be
D_{\mu}\vp \equiv (\partial_{\mu}+ieA_{\mu})\vp\ , \nonumber
\ee
\be
F_{\mu\nu} \equiv \partial_{\mu}A_{\nu}-\partial_{\nu}A_{\mu}\ , \nonumber
\ee
and the symmetry breaking (or ``Mexican hat") potential is
\be
V(\vp) = \frac{\lambda}{4}(\vp^*\vp -\eta^2)^2\ .
\ee
Here $e$ is the gauge coupling constant and $\eta$ is a parameter that
represents the scale of the symmetry breaking. In preparation for our
discussion of the more general cosmological situation, let us note
that at high temperatures the Lagrangian contains
temperature-dependent corrections given by
\be
V(\vp) \rightarrow V(\vp) + CT^2\vp^2 + \cdots \ ,
\ee
where $C$ is a constant. For $T>T_c$, where $T_c$ is defined by
\be
T_c^2 = \frac{\lambda}{2C}\eta^2\ , \nonumber
\ee
we see that
\be
V(\vp) = \frac{\lambda}{4}(\vp^*\vp) +\frac{\lambda}{2}\eta^2
\left(\frac{T^2}{T_c^2} -1\right)\vp^*\vp + \frac{\lambda}{4}\eta^2\ 
\ee
is minimized by $\langle\vp \rangle=0$. However, for $T<T_c$, minimizing the
above expression with respect to $\vp$ yields a minimum at
\be
\langle\vp \rangle^2=\eta^2\left(1-\frac{T^2}{T_c^2}\right)\ .
\ee
Thus, for $T>T_c$ the full symmetry group $G=U(1)$ is restored and the vacuum
expectation value of the $\vp$-field is zero. As the system cools, there is a
phase transition at the critical temperature $T_c$ and the vacuum symmetry
is spontaneously broken, in this case entirely:
\be
G=U(1) \longrightarrow 1=H\ .
\ee
The vacuum manifold is
\be
{\cal M} \equiv G/H= U(1)/1 =U(1)=
\{\vp\ :\ \vp=\eta e^{i\alpha}\ , \ 0\leq\alpha \leq 2\pi\}\ .
\ee
The group $U(1)$ is topologically a circle, so
\be
  {\cal M} = S^1\ .
\ee
The set of 
topologically equivalent ways to map one circle to another circle
is given by the integers.  (We can wrap it any number of times
in either sense.)  The first homotopy group is therefore the
integers.  Calculating homotopy groups in general is difficult,
but for $S^1$ all of the other groups are trivial.  We therefore have
\bea
\pi_0(S^1)&=& 1 \\
\pi_1(S^1)&=& {\cal Z} \\
\pi_2(S^1)&=& 1 \\
\pi_3(S^1)&=& 1 \ .
\eea
The Abelian Higgs model therefore allows for cosmic strings.

To see how strings form in this model, consider what happens as $T$ decreases
through $T_c$. The symmetry breaks and the field acquires a VEV given by
\be
\langle\vp \rangle =\eta e^{i\alpha}\ ,
\ee
where $\alpha$ may be chosen differently in different regions of space, as
implied by causality (we shall discuss this shortly). The requirement 
that $\langle \vp \rangle$ be single
valued implies that around any closed curve in space the change 
$\Delta\alpha$ in $\alpha$ must satisfy
\be
\Delta\alpha = 2\pi n\ , \ \ \ \ \ n\in{\cal Z}\ .
\ee
If for a given loop we have $n\neq 0$, then we can see that any
2-surface bounded by the loop must contain a singular point, for if
not then we can continuously contract the loop to a point, implying
that $n=0$ which is a contradiction. At this singular point the phase,
$\alpha$, is undefined and $\langle\vp \rangle=0$. Further,
$\langle\vp \rangle$ must be zero all along an infinite or closed
curve, since otherwise we can contract our loop without encountering a
singularity. We identify this infinite or closed curve of false vacuum
points as the core of our string.
We shall restrict our attention to the case where $n=1$, since this is the 
most likely configuration.

The first discussion of string solutions to the Abelian Higgs model is due to
Nielsen and Olesen. Assume the string is straight and that
the core is 
aligned with the $z$-axis. In cylindrical polar coordinates, $(r,\theta)$, let
us make the ansatz

\begin{eqnarray}
\vp & = & \eta X(r)e^{i\theta} \ , \nonumber \\
A_{\mu} & = & \frac{1}{e}Y(r)\partial_{\mu}\theta\ .
\end{eqnarray}
The Lagrangian then yields the simplified equations of motion
\begin{eqnarray}
-X'' -\frac{X'}{r} +\frac{Y^2X}{r^2} +\frac{\lambda\eta^2}{2}X(X^2-1) & = & 0 
\ ,\nonumber \\
-Y'' +\frac{Y'}{r} + \frac{2e^2\eta^2 X^2Y}{\lambda} & = & 0\ ,
\end{eqnarray}
where a prime denotes differentiation with respect to r.
It is not possible to solve these equations analytically, but asymptotically
we have
\begin{eqnarray}
X\sim r\ & , & Y=1 + {\cal O}(r^2)\ ; \ {\rm as }\  r\rightarrow 0\ , \nonumber \\
X\sim 1-{\cal O}\left(\frac{e^{-\lambda^{1/2}\eta r}}{\sqrt{\eta r}}\right)
\ & , &Y\sim 
{\cal O}(\sqrt{\eta r}
e^{-2^{1/2}e\eta r})\ ; \  {\rm as }\  r\rightarrow \infty\ .
\end{eqnarray}
This solution corresponds to a string
centered on the $z$-axis, with a central magnetic core of width 
$\sim (e\eta)^{-1}$ carrying a total magnetic flux
\be
\Phi = \oint_{r=\infty} A_{\mu}dx^{\mu} = -\frac{2\pi}{e}\ . \nonumber
\ee
The core region over which the Higgs fields are appreciably non-zero has
width $\sim (\lambda^{1/2}\eta)^{-1}$.  Note that these properties
depend crucially on the fact that we started with a gauged symmetry.
A spontaneously broken global symmetry with nontrivial $\pi_i(\M)$ would
still produce cosmic strings, but they would be much less localized,
since there would be no gauge fields to cancel the scalar gradients
at large distances.

Strings are characterized by their {\it tension}, which is the energy
per unit length.  For the Nielsen-Olesen solution just discussed, the
tension is approximately
\be
  \mu \sim \eta^2\ .
\ee
Thus, the energy of the string is set by the expectation value of
the order parameter responsible for the symmetry breaking.  This
behavior is similarly characteristic of other kinds of topological
defects.  Again, global defects are quite different; the tension of
a global string actually diverges, due to the slow fall-off of the
energy density as we move away from the string core.  It is often
convenient to parameterize the tension by the dimensionless
quantity $G\mu$, so that a Planck-scale string would have
$G\mu \sim 1$.

From the above discussion we can see that cosmology provides us with a
unique opportunity to explore the rich and complex structure of
particle physics theories. Although the topological solutions discussed
above exist in the theory at zero temperature, there is no mechanism 
within the theory to produce these objects. The topological structures
contribute a set of zero measure in the phase space of possible solutions to
the theory and hence the probability of production in particle 
processes is exponentially suppressed. What the cosmological evolution
of the vacuum supplies is a concrete causal mechanism for producing
these long lived exotic solutions.

At this point it is appropriate to discuss how many of these defects
we expect to be produced at a cosmological phase transition. In the
cosmological context, the mechanism for the production of defects is
known as {\it the Kibble mechanism}. The guiding 
principle here is causality. As the phase transition takes place, the
maximum causal distance imposed on the theory by cosmology is simply
the Hubble distance -- the distance which light can have traveled
since the big bang. 

As we remarked above, as the temperature of the universe falls well below the critical temperature of the phase transition, $T_c$, the expectation value
of the order parameter takes on a definite value ($\sim \eta e^{i\alpha}$ in
the Abelian Higgs model) in each region of space. However, at temperatures
around the critical temperature we expect that thermal fluctuations in
$\langle\phi \rangle$ will be large so that as the universe cools it will 
split into domains with different values of $\alpha$ in different domains.
This is the crucial role played by the cosmological evolution. 

The Kibble mechanism provides us with an order of magnitude upper
bound for the size of such a region as the causal horizon size at the time
of the phase transition. The boundaries between the domains will be regions
where the phase of $\langle \phi \rangle$ changes smoothly. If the phase
changes by $2\pi n$ for some $n\neq 0$ when traversing a loop in space, then
any surface bounded by that loop intersects a cosmic string. These strings
must be horizon-sized or closed loops. Numerical simulations of cosmic string
formation indicate that the initial distribution of strings consists of $80\%$
horizon-sized and $20\%$ loops by mass. If we
consider the temperature at which there is insufficient thermal energy to 
excite a correlation volume back into the unbroken state, the {\it Ginsburg 
temperature}, $T_G$, then, given the assumption of thermal equilibrium
above the phase transition,  a much improved estimate for the initial 
separation of the defects can be derived and is
given by
\begin{eqnarray}
\xi(t_G) \sim \lambda^{-1} \eta^{-1} \ , \nonumber
\end{eqnarray}
where $\lambda$ is the self coupling of the order parameter. This separation
is microscopic.

Interesting bounds on the tension of cosmic strings produced by
the Kibble mechanism arise from two sources:  perturbations of the CMB,
and gravitational waves.  Both arise because the motions of heavy
strings moving at relativistic velocities lead to time-dependent
gravitational fields.  The actual values of the bounds are controversial,
simply because it is difficult to accurately model the nonlinear
evolution of a string network, and the results can be sensitive to
what assumptions are made.  Nevertheless, the CMB bounds amount roughly
to \cite{Landriau:2003xf,Pogosian:2003mz,Vincent:1997cx,Moore:2001px}
\be
  G\mu \leq 10^{-6}\ .
\ee
This corresponds roughly to strings at the GUT scale, $10^{16}$~GeV,
which is certainly an interesting value.
Bounds from gravitational waves come from two different techniques:
direct observation, and indirect measurement through accurate pulsar
timings.  Currently, pulsar timing measurements are more constraining,
and give a bound similar to that from the CMB 
\cite{Caldwell:1996en}. 
Unlike the CMB measurements, however, gravitational wave observatories
will become dramatically better in the near future, through
operations of ground-based observatories such as LIGO as well as
satellites such as LISA.  These experiments should be able to improve
the bounds on $G\mu$ by several orders of magnitude 
\cite{Damour:2001bk}.

Now that we have established the criteria necessary for the
production of topological defects in spontaneously broken theories, let us 
apply these conditions to the best understood physical example, the electroweak
phase transition. As we remarked earlier, the GWS theory is based on the
gauge group $SU(2)_L\times U(1)_Y$. At the phase transition this breaks to
pure electromagnetism, $U(1)_{em}$. Thus, the vacuum manifold is the space
of cosets
\be
{\cal M}_{EW} = [SU(2)\times U(1)]/U(1) \ ,
\ee
This looks complicated but in fact this space is topologically equivalent
to the three-sphere, $S^3$.  The homotopy groups of the three-sphere
are
\bea
\pi_0(S^3)&=& 1 \\
\pi_1(S^3)&=& 1 \\
\pi_2(S^3)&=& 1 \\
\pi_3(S^3)&=& {\cal Z} \ .
\eea
(Don't be fooled into thinking that all homotopy groups of spheres
vanish except in the dimensionality of the sphere itself; for
example, $\pi_3(S^2)={\cal Z}$.)

Thus, the electroweak model does not lead to walls, strings, or
monopoles.  It does lead to what we called ``texture,'' which 
deserves further comment.  In a theory where $\pi_3({\cal M})$
is nontrivial but the other groups vanish, 
we can always map three-dimensional space smoothly
into the vacuum manifold; there will not be a defect where the
field climbs out of ${\cal M}$.  However, if we consider field
configurations which approach a unique value at spatial infinity,
they will fall into homotopy classes characterized by elements
of $\pi_3({\cal M})$; configurations with nonzero winding will be
textures.  If the symmetry is global, such configurations will 
necessarily contain gradient energies from the scalar fields.  The
energy perturbations caused by global textures were, along with
cosmic strings, formerly popular as a possible origin of structure
formation in the universe \cite{spertur1,spertur2}; the predictions of these
theories are inconsistent with the sharp acoustic peaks observed in
the CMB, so such models are no longer considered viable.

In the standard model, however, the broken symmetry is gauged.  In
this case there is no need for gradient energies, since the gauge 
field can always be chosen to cancel them; equivalently, 
``texture'' configurations can always be brought to the vacuum by
a gauge transformation.  However, {\it transitions} from one 
texture configuration to one with a different winding number are
gauge invariant.  These transitions will play a role in 
electroweak baryon number violation, discussed in the next section.

\subsection{Baryogenesis}

The symmetry between particles and antiparticles \cite{Dirac1,Dirac2}, firmly
established in collider physics, naturally leads
to the question of why the observed universe 
is composed almost entirely of matter with little or no primordial antimatter. 

Outside of particle accelerators, antimatter can be seen in cosmic rays
in the form of a few antiprotons, present at a
level of around $10^{-4}$ in comparison with the number of protons
(for example see \cite{SA 88}). 
However, this proportion is 
consistent with secondary antiproton production through accelerator-like
processes, $p+p\rightarrow 3p + {\bar p}$, as the cosmic rays stream
towards us. Thus there is no evidence for primordial antimatter in 
our galaxy. Also, if matter and antimatter galaxies were to coexist
in clusters of galaxies, then we would expect there to be a detectable
background of $\gamma$-radiation from nucleon-antinucleon annihilations 
within
the clusters. This background is not observed and so we conclude that
there is negligible antimatter on the scale of clusters (For a review of the 
evidence for a baryon asymmetry see \cite{GS 76}.)

More generally, if large domains of matter and
antimatter exist, then annihilations would take place at the 
interfaces between them. If the typical size of such a domain was small
enough, then the energy released by these annihilations would
result in a diffuse $\gamma$-ray background and a distortion of the cosmic 
microwave radiation, neither of which is observed.

While the above considerations put an experimental upper bound on the
amount of antimatter in the universe, strict quantitative estimates of
the relative abundances of baryonic matter and antimatter may also be
obtained from the standard cosmology.  The baryon number density does
not remain constant during the evolution of the universe, instead
scaling like $a^{-3}$, where $a$ is the cosmological scale factor
\cite{book}. It is therefore convenient to define the baryon asymmetry
of the universe in terms of the quantity 
\be
 \eta\equiv \frac{n_B}{s} \ , 
\ee 
defined earlier.  Recall that the range of
$\eta$ consistent with the deuterium and $^3$He primordial abundances
is
\begin{equation}
2.6\times 10^{-10} < \eta
 < 6.2\times 10^{-10} \ .
\end{equation}
Thus the natural question arises; as the universe cooled from early
times to today, what processes, both particle physics and
cosmological, were responsible for the generation of this very
specific baryon asymmetry? (For reviews of mechanisms to generate the baryon asymmetry, see

As pointed out by Sakharov \cite{sak}, a small baryon asymmetry $\eta$ 
may have been produced in the early universe if three necessary conditions 
are satisfied
\begin{itemize}
\item baryon number ($B$) violation,
\item violation of 
$C$ (charge conjugation symmetry) and $CP$ (the composition of parity and 
$C$),
\item departure from thermal equilibrium. 
\end{itemize}

The first 
condition should be clear since, starting from a  baryon symmetric universe 
with $\eta=0$, baryon number violation must take place in order to  evolve 
into  a universe in which $\eta$ does not vanish. The second Sakharov 
criterion is required 
because, if $C$ and $CP$ are exact symmetries, one can prove
that the total rate for any process which produces an excess of baryons is
equal to the rate of the complementary process which produces an
excess of antibaryons and so no net baryon number can be created. 
That is to say that the thermal average of the baryon number operator $B$, 
which is odd under
both $C$ and $CP$, is zero unless those discrete symmetries are
violated. $CP$ violation is present either if  there are complex phases in 
the Lagrangian which cannot be reabsorbed by field redefinitions (explicit 
breaking) or if some Higgs scalar field acquires a VEV which is not real 
(spontaneous breaking). We will discuss this in detail shortly.

Finally, to explain the third criterion,  one can  calculate the equilibrium 
average of $B$ at a temperature $T=1/\beta$:
\bea
\langle B\rangle_T & = & {\rm Tr}\:(e^{-\beta H}B)=
 {\rm Tr}\:[(CPT)(CPT)^{-1}e^{-\beta H}B)] \nonumber \\
& = & {\rm Tr}\:(e^{-\beta H}(CPT)^{-1}B(CPT)]=
-{\rm Tr}\:(e^{-\beta H}B) \ ,
\eea
where we  have used that the Hamiltonian $H$ commutes with $CPT$. Thus
$\langle B\rangle_T = 0$ in equilibrium and there is no generation of
net baryon number.

Of the three Sakharov conditions, baryon number violation and $C$ and 
$CP$ violation may be investigated  only within a given particle physics 
model, while the third condition -- the departure from thermal equilibrium -- 
may be discussed in a more general way, as we shall see (for baryogenesis reviews 
see~\cite{AD 92,riotto 98,Trodden:1998ym,Riotto:1999yt,review,NTreview,CKNreview}.)  
Let us discuss the Sakharov criteria in more detail.

\subsection{Baryon Number Violation}

\subsubsection{$B$-violation in Grand Unified Theories}

As discussed earlier,
Grand Unified Theories (GUTs) \cite{lan} describe the fundamental 
interactions by means of a unique gauge group $G$ which contains the 
Standard Model (SM) gauge group $SU(3)_C\otimes SU(2)_L\otimes U(1)_Y$. 
The fundamental idea of GUTs is that at energies higher than a certain 
energy threshold $M_{{\rm GUT}}$ the group symmetry is $G$ and that, at 
lower energies, the symmetry is broken down to the SM gauge symmetry, 
possibly through a chain of symmetry breakings. 
The main motivation for this scenario is that, at least in supersymmetric
models, the (running) gauge 
couplings of the SM unify  \cite{couplingunity,n2,n3} at the scale $
M_{{\rm GUT}} \simeq 2\times 10^{16}$ GeV, hinting at the presence 
of a GUT involving a higher symmetry with a single gauge coupling.

Baryon number violation seems very natural in GUTs. Indeed, 
a general property of these theories is that the same representation of 
$G$ may contain both 
quarks and leptons, and therefore it is possible for scalar and gauge bosons 
to mediate gauge interactions among fermions having different baryon 
number.

\subsubsection{$B$-violation in the Electroweak theory.}

It is well-known that the most general 
renormalizable Lagrangian invariant 
under the SM gauge group and containing only color singlet Higgs fields 
is automatically invariant under global abelian 
symmetries which may be identified with the baryonic and leptonic symmetries. 
These, therefore, are accidental symmetries and as a result it is not 
possible to violate $B$ and $L$ at tree-level or at any order of 
perturbation theory. Nevertheless, in many cases the perturbative expansion 
does not describe all the dynamics of the theory and, indeed, in 1976 't 
Hooft \cite{ho}
realized that nonperturbative effects (instantons) may give rise to processes 
which violate the combination $B+L$, but not the orthogonal combination 
$B-L$. The probability of these processes occurring today is exponentially 
suppressed and probably irrelevant. However, in more extreme situations -- 
like the primordial universe at very high temperatures 
\cite{ds,manton,km,krs} -- baryon and lepton number violating processes 
may be fast enough to play a significant role in baryogenesis. Let us have 
a closer look.

At the quantum level, the baryon and the lepton symmetries are  
anomalous \cite{adler,bj}, so that their respective Noether currents $j_B^{\mu}$ and $j_L^{\mu}$ are
no longer conserved, but satisfy
\be
\partial_{\mu}j_B^{\mu} = \partial_{\mu}j_L^{\mu}
=n_f\left(\frac{g^2}{32\pi^2}W_{\mu\nu}^a {\tilde W}^{a\mu\nu}
-\frac{g'^2}{32\pi^2}F_{\mu\nu}{\tilde F}^{\mu\nu}\right) \ ,
\label{anomaly}
\ee
where $g$ and $g'$ are the gauge couplings of $SU(2)_L$ and $U(1)_Y$, 
respectively,  $n_f$ is the number of families and $
\tilde{W}^{\mu\nu} = (1/2) \epsilon^{\mu\nu\alpha\beta}W_{\alpha\beta}$ 
is the dual of the $SU(2)_L$ field strength tensor, with an analogous
expression holding for ${\tilde F}$. To understand how the anomaly is closely 
related to the vacuum structure of the theory, we may compute  the change in
baryon number from time $t=0$ to some arbitrary final time $t=t_f$.
For transitions between vacua, the average values of the field strengths 
are zero at the beginning and the end of the evolution. The change in baryon 
number may be written as
\be
\Delta B = \Delta N_{CS} \equiv n_f[N_{CS}(t_f) - N_{CS}(0)]\ .
\ee
where the Chern-Simons number is defined to be  
\begin{equation}
N_{CS}(t) \equiv \frac{g^2}{32 \pi^2}\int d^3x\, \epsilon^{ijk}
\:{\rm Tr}\:\left( A_i \partial_j A_k + \frac{2}{3}ig A_i A_j A_k \right)\ ,
\label{NCSdef}
\end{equation}
where $A_i$ is the SU(2)$_L$ gauge field.
Although the Chern-Simons number is not gauge invariant, the change 
$\Delta N_{CS}$ is. Thus, changes in Chern-Simons number result in 
changes in baryon number which are integral multiples of the number of
families $n_f$ (with $n_f=3$ in the real world). 
Gauge transformations $U(x)$ which connects two degenerate 
vacua of the gauge theory  may  change the Chern-Simons number by an integer 
$n$, the winding number. 

If the system is able to perform a transition from 
the vacuum ${\cal G}_{{\rm vac}}^{(n)}$ to the closest one
${\cal G}_{{\rm vac}}^{(n\pm 1)}$, the Chern-Simons number is changed by 
unity and $\Delta B=\Delta L=n_f$. 
Each transition creates 9 left-handed quarks (3 color states for each 
generation) and 3 left-handed leptons (one per generation). 

However, adjacent vacua of the electroweak theory are separated by a ridge of
configurations with energies larger than that of the vacuum.  
The lowest energy point on this ridge is a saddle point solution
to the equations of motion with a single negative eigenvalue, and
is referred to as the {\it sphaleron} \cite{manton,km}.
The probability of baryon number nonconserving processes at zero temperature
has been computed by 't Hooft \cite{ho} and is highly suppressed. 

The thermal rate
of baryon number violation in the {\it broken} phase is \be
\Gamma_{sp}(T) = \mu\left(\frac{M_W}{\alpha_W T}\right)^3M_W^4 
\exp\left(-\frac{E_{sph}(T)}{T}\right) \ ,
\label{brokenrate}
\ee
where $\mu$ is a dimensionless constant.
Although the Boltzmann suppression in~(\ref{brokenrate}) appears large,
it is to 
be expected that, when the electroweak symmetry becomes restored  at a 
temperature of around $100\,$GeV, there will no longer be an exponential 
suppression factor. A simple 
estimate is that the rate per unit volume of sphaleron events is
\be
\Gamma_{sp}(T)=\kappa(\alpha_W T)^4 \ ,
\label{unbrokenrate}
\ee
with $\kappa$ another dimensionless constant. The rate of sphaleron processes
can be related to the diffusion constant for Chern-Simons 
number  by a fluctuation-dissipation theorem
\cite{ks88} (for a good description of this see \cite{review}).

\subsubsection{CP violation}

$CP$ violation in GUTs arises in loop-diagram corrections to 
baryon number violating bosonic decays. Since it is necessary that 
the particles in the loop also undergo 
$B$-violating decays, the relevant particles are the $X$, $Y$, and 
$H_3$ bosons in the case of $SU(5)$. 

In the electroweak theory things are somewhat different.
Since  only the left-handed 
fermions are $SU(2)_L$ gauge coupled, $C$ is maximally broken in the SM.  
Moreover, $CP$ is known not to be an exact symmetry
of the weak interactions. This is seen experimentally in the neutral 
kaon system through $K_0$, ${\bar K}_0$ mixing. Thus, 
$CP$ violation is a natural feature of the
standard electroweak model. 

While this is encouraging for baryogenesis, it turns out that this
particular source of $CP$ violation is not strong enough. The relevant
effects are parameterized by a dimensionless constant which is no
larger than $10^{-20}$. This appears to be much too small to account
for the observed BAU and, thus far, attempts to utilize this source of
CP violation for electroweak baryogenesis have been unsuccessful. In
light of this, it is usual to extend the SM in some fashion
that increases the amount of $CP$ violation in the theory while not
leading to results that conflict with current experimental data. One
concrete example of a well-motivated extension in the minimal
supersymmetric standard model (MSSM).

\subsubsection{Departure from Thermal Equilibrium}

In some scenarios, such as GUT baryogenesis, the third Sakharov
condition is satisfied due to the presence of superheavy decaying
particles in a rapidly expanding universe. These generically fall
under the name of out-of-equilibrium decay mechanisms.

The underlying idea is fairly simple. 
If the decay rate $\Gamma_X$ of the superheavy particles $X$ 
at the time they become 
nonrelativistic ({\it i.e.} at the temperature $T\sim M_X$) is much smaller 
than the expansion rate of the universe, then the $X$  
particles cannot decay on the time scale of the expansion and so they 
remain as 
abundant as photons for $T\lsim M_X$. In other words, at some  temperature 
$T > M_X$, the superheavy particles  
$X$ are so weakly interacting that they cannot catch up with the expansion 
of the universe and they decouple from the thermal bath while they are still 
relativistic, so that $n_{X}\sim n_\gamma\sim T^3$ at the time of decoupling. 

Therefore, at temperature $T\simeq M_X$, they populate the universe 
with an abundance which is much larger than the equilibrium one. 
This overabundance is precisely 
the departure from thermal equilibrium needed to produce a final nonvanishing 
baryon asymmetry when the heavy states $X$ undergo $B$ and $CP$ violating
decays.

The out-of-equilibrium condition requires very heavy states: $
M_X\gsim (10^{15}-10^{16})\:{\rm GeV}$ and $M_X\gsim
(10^{10}-10^{16})\:{\rm GeV}$, for gauge and scalar bosons,
respectively \cite{decay}, if these heavy particles decay through
renormalizable operators.

A different implementation can be found in the electroweak theory. At
temperatures around the electroweak scale, the expansion rate of the
universe in thermal units is small compared to the rate of baryon
number violating processes. This means that the equilibrium
description of particle phenomena is extremely accurate at electroweak
temperatures. Thus, baryogenesis cannot occur at such low scales
without the aid of phase transitions and the question of the order of
the electroweak phase transition becomes central.

If the EWPT is second order or a continuous crossover, the associated
departure from equilibrium is insufficient to lead to relevant baryon
number production \cite{krs}.  This means that for EWBG to succeed, we
either need the EWPT to be strongly first order or other methods of
destroying thermal equilibrium to be present at the phase transition.
For a first order transition there is an extremum at
$\vp=0$ which becomes separated from a second local minimum by an
energy barrier.  At the critical temperature $T=T_c$ both phases are
equally favored energetically and at later times the minimum at $\vp
\neq 0$ becomes the global minimum of the theory.

The dynamics of the phase transition in this situation is crucial to
most scenarios of electroweak baryogenesis. The essential picture is
that around $T_c$ quantum tunneling occurs and nucleation of bubbles
of the true vacuum in the sea of false begins. Initially these bubbles
are not large enough for their volume energy to overcome the competing
surface tension and they shrink and disappear. However, at a
particular temperature below $T_c$, bubbles just large enough to grow
nucleate. These are termed {\it critical} bubbles, and they expand,
eventually filling all of space and completing the transition.

As the bubble walls
pass each point in space, the order
parameter changes rapidly, as do the other fields, and this leads to a
significant departure from thermal equilibrium. Thus, if the phase 
transition is strongly enough first order it is possible to satisfy
the third Sakharov criterion in this way.

A further natural way to depart from equilibrium is provided by the
dynamics of topological defects~\cite{DMEWBGii,DMEWBGiii,rob-annei,rob-anneii}. 
If, for example, cosmic strings are
produced at the GUT phase transition, then the decays of loops of
string act as an additional source of superheavy bosons which undergo
baryon number violating decays.

When defects are produced at the TeV scale, a detailed analysis of the
dynamics of a network of these objects shows that a significant baryon
to entropy ratio can be generated if the electroweak symmetry is
restored around such a higher scale ordinary defect.
Although $B$-violation can be inefficient along nonsuperconducting strings \cite{cemr}, there remain
viable scenarios involving other ordinary defects, superconducting strings 
or defects carrying  baryon number.

\subsubsection{Baryogenesis via leptogenesis}

Since the linear combination $B-L$ is left unchanged by sphaleron
transitions, the baryon asymmetry may be generated from a lepton
asymmetry \cite{goran,fy,gelmann} (see also~\cite{luty1,luty}.)  Indeed, sphaleron transition will reprocess any
lepton asymmetry and convert (a fraction of) it into baryon
number. This is because $B+L$ must be vanishing and the final baryon
asymmetry results to be $B\simeq -L$.

In the SM as well as in its unified extension based on the group
$SU(5)$, $B-L$ is conserved and no asymmetry in $B-L$ can be
generated.  However, adding right-handed Majorana neutrinos to the SM
breaks $B-L$ and the primordial lepton asymmetry may be generated by
the out-of-equilibrium decay of heavy right-handed Majorana neutrinos
$N_L^c$ (in the supersymmetric version, heavy scalar neutrino decays
are also relevant for leptogenesis).  This simple extension of the SM
can be embedded into GUTs with gauge groups containing $SO(10)$. Heavy
right-handed Majorana neutrinos can also explain the smallness of the
light neutrino masses via the see-saw mechanism \cite{gelmann}. 

\subsubsection{Affleck-Dine Baryogenesis}
Finally in this section, we mention briefly a mechanism introduced by Affleck 
and Dine~\cite{affleckdine} involving the cosmological evolution of scalar fields 
carrying baryonic charge.

Consider a colorless, electrically neutral combination of quark and
lepton fields. In a supersymmetric theory this object has a scalar 
superpartner, $\chi$, composed of the corresponding squark ${\tilde q}$ 
and slepton ${\tilde l}$ fields. 

An important feature of supersymmetric field 
theories is the existence of ``flat directions" in field space, on which 
the scalar potential vanishes. Consider the case where some component of 
the field $\chi$ lies along a flat direction. By this we mean that there exist
directions in the superpotential along which the relevant components of 
$\chi$ can be considered as a free massless field.
At the level of renormalizable terms, flat directions are generic, but
supersymmetry breaking and nonrenormalizable operators lift the flat 
directions and sets the scale for  their potential.

During inflation it is likely that the $\chi$ field is displaced from the 
position $\langle\chi\rangle=0$, establishing the initial conditions for 
the subsequent evolution of the field. An important role is played at this 
stage by baryon number violating operators in the potential $V(\chi)$, which 
determine the initial phase of the field. When the Hubble rate becomes of the 
order of the curvature of the potential $\sim m_{3/2}$, the condensate starts 
oscillating around its present minimum. At this time, $B$-violating terms in 
the potential are of comparable importance to the mass term, thereby 
imparting a substantial baryon number to the condensate. After this 
time, the baryon number violating operators are negligible so that, when the 
baryonic charge of $\chi$ 
is transferred to fermions through decays, the  net baryon number of the 
universe is  preserved by the subsequent cosmological evolution. 

The most recent implementations of the Affleck-Dine scenario have been in
the context of the minimal supersymmetric standard model \cite{Lisa1,Lisa2}
in which, because there are large numbers of fields, flat directions occur
because of accidental degeneracies in field space. 

\section{Inflation}

In our previous lectures we have described what is known as {\it
the standard cosmology}. This framework is a towering achievement,
describing to great accuracy the physical processes leading to the
present day universe. However, there remain outstanding issues in
cosmology. Many of these come under the heading of initial condition
problems and require a more complete description of the sources of
energy density in the universe. The most severe of these problems
eventually led to a radical new picture of the physics of the early
universe - {\it cosmological inflation} \cite{guthi,lindei,asi}, 
which is the subject of this lecture.

We will begin by describing some of the problems of the
standard cosmology.

\subsection{The Flatness Problem}
The Friedmann equation may be written as
\begin{equation}
\label{alternatefriedmann}
\Omega -1=\frac{k}{H^2a^2} \ ,
\end{equation}
where for brevity we are now writing $\Omega$ instead of $\Omega_{\rm
total}$.  Differentiating this with respect to the scale factor, this implies
\begin{equation}
\label{omegaofa}
\frac{d\Omega}{da}=(1+3w)\frac{\Omega(\Omega -1)}{a} \ .
\end{equation}
This equation is easily solved, but its most general properties are
all that we shall need and they are qualitatively different depending
on the sign of $1+3w$. There are three fixed points of this
differential equation, as given in table~4.
\begin{table}[ht]
\begin{center}
\begin{tabular}{l|l|l}
Fixed Point & $1+3w>0$ & $1+3w<0$ \\ \hline
$\Omega=0$ & attractor & repeller \\
$\Omega=1$ & repeller & attractor \\
$\Omega=\infty$ & attractor & repeller \\
\end{tabular}
\end{center}
\caption{Behavior of the density parameter near fixed points.}
\end{table}

Observationally we know that $\Omega \simeq 1$ today -- {\it i.e.}, we are
very close to the repeller of this differential equation for a universe
dominated by ordinary matter and radiation ($w > -1/3$). Even if we
only took account of the luminous matter in the universe, we would
clearly live in a universe that was far from the attractor points of
the equation. It is already quite puzzling that the universe has not
reached one of its attractor points, given that the universe has
evolved for such a long time. However, we may be more quantitative
about this. If the only matter in the universe is radiation and dust,
then in order to have $\Omega$ in the range observed today requires
(conservatively)
\begin{equation}
\label{omegaconstraints}
0\leq 1-\Omega \leq 10^{-60} \ .
\end{equation}
This remarkable degree of fine tuning 
is the flatness problem. Within the context of the standard cosmology
there is no known explanation of this fine-tuning.

\subsection{The Horizon Problem}

\begin{figure}
  \centerline{
  \psfig{figure=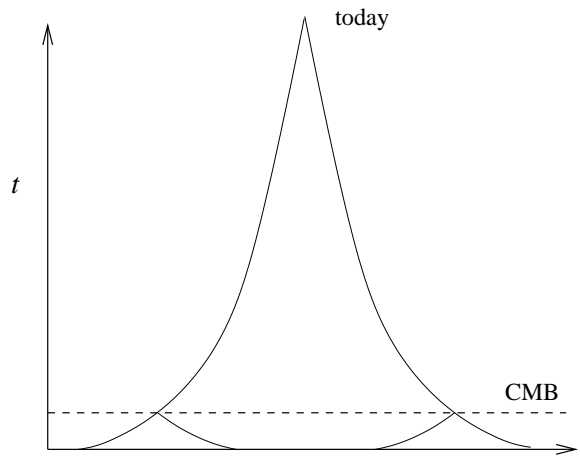,angle=0,height=6cm}}
  \caption{Past light cones in a universe expanding from a Big
  Bang singularity, illustrating particle horizons in cosmology.
  Points at recombination, observed today as parts of the cosmic
  microwave background on opposite sides of the sky, have 
  non-overlapping past light cones (in conventional cosmology);
  no causal signal could have influenced them to have the same
  temperature.}
  \label{horizonfig}
\end{figure}

The {\it horizon problem}
stems from the existence of particle horizons in
FRW cosmologies, as discussed in the first lecture.  
Horizons exist because there is only a
finite amount of time since the Big Bang singularity, and
thus only a finite distance that photons can travel within
the age of the universe.  Consider a photon moving along
a radial trajectory in a flat universe (the
generalization to non-flat universes is straightforward).
In a flat universe, we can normalize the scale factor to
\be
  a_0 = 1
\ee
without loss of generality.
A radial null path obeys
\be
  0 = ds^2 = -dt^2 + a^2 dr^2\ ,
  \label{8.501}
\ee
so the comoving (coordinate) distance traveled by such a photon 
between times $t_1$ and $t_2$ is
\be
  \Delta r = \int^{t_2}_{t_1} {{dt}\over{a(t)}}\ .
  \label{8.502}
\ee
To get the physical distance as it would be measured by an
observer at any time $t$, simply multiply by $a(t)$.
For simplicity let's imagine we are in a matter-dominated universe,
for which 
\be
  a = \left({t\over t_0}\right)^{2/3}\ .
  \label{8.550}
\ee
The Hubble parameter is therefore given by
\bea
  H &=& {2 \over 3} t^{-1} \cr 
  &=& a^{-3/2}H_0 \ .
  \label{8.551}
\eea
Then the photon travels a comoving distance
\be
  \Delta r = 2H_0^{-1}\left(\sqrt{a_2} - \sqrt{a_1}\right) \ .
  \label{8.503}
\ee
The comoving horizon size when $a=a_*$ is the distance a photon
travels since the Big Bang,
\be
  r_{\rm hor}(a_*)= 2H_0^{-1}\sqrt{a_*}\ .
  \label{8.552}
\ee
The physical horizon size, as measured on the spatial
hypersurface at $a_*$, is therefore simply
\be
  d_{\rm hor}(a_*)= a_* r_{\rm hor}(a_*) = 2H_*^{-1}\ .
  \label{8.553}
\ee
Indeed, for any nearly-flat universe containing a mixture of
matter and radiation, at any one epoch we will have
\be
  d_{\rm hor}(a_*) \sim H_*^{-1}\ ,
  \label{8.554}
\ee
where $H_*^{-1}$ is the Hubble distance at that particular epoch.
This approximate equality leads to a strong temptation to use
the terms ``horizon distance'' and ``Hubble distance'' 
interchangeably; this temptation should be resisted, since inflation
can render the former much larger than the latter, as we will soon
demonstrate.  

The horizon
problem is simply the fact that the CMB is isotropic to a high
degree of precision, even though widely separated points on the
last scattering surface are completely outside each others'
horizons.  When we look at the CMB we were observing the universe
at a scale factor $a_{\rm CMB}\approx 1/1200$; meanwhile, the
comoving distance between a point on the CMB and an observer on Earth is
\bea
  \Delta r &=& 2H_0^{-1}\left(1-\sqrt{a_{\rm CMB}}\right)\cr
  &\approx& 2H_0^{-1} \ .
  \label{8.555}
\eea
However, the comoving horizon distance for such a point is
\bea
  r_{\rm hor}(a_{\rm CMB})&=& 2H_0^{-1}\sqrt{a_{\rm CMB}}\cr
  &\approx& 6\times 10^{-2} H_0^{-1}\ .
  \label{8.556}
\eea
Hence, if we observe two widely-separated parts of the CMB, they
will have non-overlapping horizons;
distinct patches of the CMB sky were
causally disconnected at recombination.  Nevertheless, they
are observed to be at the same temperature to high precision.
The question then is, how did they know ahead of time to
coordinate their evolution in the right way, even though they
were never in causal contact?  We must somehow modify the
causal structure of the conventional FRW cosmology.

\subsection{Unwanted Relics}

We have spoken in the last lecture about grand unified theories (GUTs),
and also about topological defects.  If grand unification occurs
with a simple gauge group $G$, any spontaneous breaking of $G$
satisfies $\pi_2(G/H)=\pi_1(H)$ for any simple subgroup $H$. In particular, 
breaking down to the standard model will lead to magnetic monopoles, since
\be
  \pi_2(G/H)=\pi_1([\su3\times\su2\times\u1]/{\cal Z}_6)={\cal Z}\ .
\ee
(The gauge group of the standard model
is, strictly speaking, $[\su3\times\su2\times\u1]/{\cal Z}_6$.  The
${\cal Z}_6$ factor, with which you may not be familiar, only affects the
global structure of the group and not the Lie algebra, and thus
is usually ignored by particle physicists.)

Using the Kibble mechanism, the expected relic abundance of 
monopoles works out to be
\be
  \Omega_{0, {\rm mono}} \sim 10^{11} \left({{T_{\rm GUT}}\over
  {10^{14}~{\rm GeV}}}\right)^3 \left({{m_{\rm mono}}\over
  {10^{16}~{\rm GeV}}}\right)\ .
\ee
This is far too big; the monopole abundance in GUTs is a
serious problem for cosmology if GUTs have anything to do with
reality.

In addition to monopoles, there may be other model-dependent
relics predicted by your favorite theory.  If these are incompatible
with current limits, it is necessary to find some way to dilute
their density in the early universe.

\subsection{The General Idea of Inflation}
The horizon problem especially
is an extremely serious problem for the standard
cosmology because at its heart is simply causality. Any solution to this
problem is therefore almost certain to require an important
modification to how information can propagate in the early
universe. Cosmological inflation is such a mechanism.

Before getting into the details of inflation we will just sketch the
general idea here. The fundamental idea is that the universe undergoes
a period of accelerated expansion, defined as a period when ${\ddot a}>0$, at early times. 
The effect of this acceleration is
to quickly expand a small region of space to a huge size, diminishing spatial curvature in the
process, making the universe 
extremely close to flat.  In addition, the horizon size is greatly
increased, so that distant points on the CMB
actually are in causal contact and unwanted relics are tremendously
diluted, solving the monopole problem.  As an unexpected bonus,
quantum fluctuations make it impossible for inflation to smooth out
the universe with perfect precision, so there is a spectrum of remnant
density perturbations; this spectrum turns out to be approximately
scale-free, in good agreement with observations of our current
universe.

\subsection{Slowly-Rolling Scalar Fields}
If inflation is to solve the problems of the standard cosmology, then
it must be active at extremely early times. Thus, we would like to
address the earliest times in the universe amenable to a classical
description. We expect this to be at or around the Planck time
$t_P$ and since Planckian quantities arise often in inflation we will 
retain values of the Planck mass in the equations of this section. 
There are {\it many} models of inflation, but because of time
constraints we will concentrate almost exclusively on the {\it
chaotic inflation} model of Linde. We have borrowed heavily in places
here from the excellent text of Liddle and Lyth \cite{ll}.

Consider modeling matter in the early universe by a real scalar field
$\phi$, with potential $V(\phi)$. The energy-momentum tensor for
$\phi$ is
\begin{equation}
\label{scalaremtensor}
T_{\mu\nu}=(\nabla_{\mu}\phi)(\nabla_{\nu}\phi)-g_{\mu\nu}\left[\frac{1}{2}g^{\alpha\beta}(\nabla_{\alpha}\phi)(\nabla_{\beta}\phi)+V(\phi)\right] \ .
\end{equation}
For simplicity we will specialize to the homogeneous case, in which
all quantities depend only on cosmological time $t$ and set $k=0$. 
A homogeneous real scalar field behaves as
a perfect fluid with
\begin{eqnarray}
\label{scalarrhoandp}
\rho_{\phi} & = & \frac{1}{2}{\dot \phi}^2 +V(\phi) \\
p_{\phi} & = & \frac{1}{2}{\dot \phi}^2 -V(\phi) \ .
\end{eqnarray}
The equation of motion for the scalar field is given by
\begin{equation}
\label{scalareom}
{\ddot \phi}+3\frac{{\dot a}}{a}{\dot \phi}+\frac{dV}{d\phi}=0 \ ,
\end{equation}
which can be thought of as the usual equation of motion for a scalar
field in Minkowski space, but with a friction term due to the
expansion of the universe.  The
Friedmann equation with such a field as the sole energy source is
\begin{equation}
\label{scalarfriedmann}
H^2= \frac{8\pi
G}{3}\left[\frac{1}{2}{\dot \phi}^2 +V(\phi)\right] \ .
\end{equation}

A very specific way in which accelerated expansion can occur is if the
universe is dominated by an energy component that approximates a
cosmological constant. In that case the associated expansion rate will
be exponential, as we have already seen. Scalar fields can accomplish
this in an interesting way.  From~(\ref{scalarrhoandp}) it is clear
that if ${\dot \phi}^2 \ll V(\phi)$ then the potential energy of the
scalar field is the dominant contribution to both the energy density
and the pressure, and the resulting equation of state is $p \simeq
-\rho$, approximately that of a cosmological constant. the resulting
expansion is certainly accelerating. In a loose sense, this negligible
kinetic energy is equivalent to the fields slowly rolling down its
potential; an approximation which we will now make more formal.

Technically, the {\it slow-roll approximation} for inflation involves
neglecting the ${\ddot \phi}$ term in~(\ref{scalareom}) and neglecting
the kinetic energy of $\phi$ compared to the potential energy. The
scalar field equation of motion and the Friedmann equation then become
\begin{equation}
\label{scalarslowroll}
{\dot \phi} \simeq -\frac{V'(\phi)}{3H} \ ,
\end{equation}
\begin{equation}
\label{friedmannsllowroll}
H^2 \simeq \frac{8\pi G}{3} V(\phi) \ ,
\end{equation}
where in this lecture a prime denotes a derivative with respect to $\phi$.

These conditions will hold if the two {\it slow-roll conditions} are
satisfied. These are
\begin{eqnarray}
|\epsilon| & \ll & 1 \nonumber \\
|\eta| & \ll & 1 \ ,
\end{eqnarray}
where the {\it slow-roll parameters} are given by
\begin{equation}
\label{epsilonslowroll}
\epsilon\equiv \frac{\mpl^2}{2}\left(\frac{V'}{V}\right)^2 \ ,
\end{equation}
and
\begin{equation}
\label{etaslowroll}
\eta\equiv \mpl^2 \frac{V''}{V} \ .
\end{equation}

It is easy to see that the slow roll conditions yield
inflation. Recall that inflation is defined by ${\ddot a}/a>0$. We can
write
\begin{equation}
\frac{\ddot a}{a}={\dot H}+H^2 \ ,
\end{equation}
so that inflation occurs if
\begin{equation}
\frac{\dot H}{H^2}>-1 \ .
\end{equation}
But in slow-roll
\begin{equation}
\frac{\dot H}{H^2} \simeq -\epsilon \ ,
\end{equation}
which will be small.  Smallness of the other parameter $\eta$ helps
to ensure that inflation will continue for a sufficient period.

It is useful to have a general expression to
describe how much inflation occurs, once it has begun. This is
typically quantified by the number of {\it e-folds}, defined by
\begin{equation}
N(t)\equiv \ln\left(\frac{a(t_{\rm end})}{a(t)}\right) \ .
\end{equation}
Usually we are interested in how many efolds occur between a given
field value $\phi$ and the field value at the end of inflation
$\phi_{\rm end}$, defined by $\epsilon(\phi_{\rm end})=1$. We also
would like to express $N$ in terms of the potential. Fortunately this
is simple to do via
\begin{equation}
\label{numberofefolds}
N(t)\equiv \ln\left(\frac{a(t_{\rm end})}{a(t)}\right)
=\int_t^{t_{\rm end}}H\ dt 
\simeq \frac{1}{\mpl^2}\int_{\phi}^{\phi_{\rm end}}\frac{V}{V'}\ d\phi \ .
\end{equation} 

The issue of initial conditions for inflation is one that is quite
subtle and we will not get into a discussion of that here. Instead we
will remain focused on chaotic inflation, in which we assume that the
early universe emerges from the Planck epoch with the scalar field
taking different values in different parts of the universe, with
typically Planckian energies. There will then be some probability for
inflation to begin in some places, and we shall focus on those.

\subsection{Attractor Solutions in Inflation}
For simplicity, let us consider a particularly simple potential
\begin{equation}
\label{oscpotential}
V(\phi)=\frac{1}{2}m^2 \phi^2 \ ,
\end{equation}
where $m$ has dimensions of mass. We shall also assume initial
conditions that at the end of the quantum epoch, which we label as
$t=0$, $\rho\sim \mpl^4$.

The slow-roll conditions give that, at $t=0$,
\begin{equation}
\label{phiinitial}
\phi=\phi_0 \sim \frac{\mpl^2}{m} \sim \frac{\mpl}{\varepsilon} \ ,
\end{equation}
where we have defined $\varepsilon\equiv m/\mpl \ll 1$.
We also have
\begin{equation}
H=\alpha\phi \ ,
\end{equation}
where
\begin{equation}
\alpha\equiv \sqrt{\frac{4\pi}{3}}\varepsilon \ ,
\end{equation}
so that the scalar field equation of motion becomes
\begin{equation}
{\ddot \phi}+3\alpha\phi{\dot \phi}+m^2\phi=0 \ .
\end{equation}

This is solved by
\begin{equation}
\label{attractorsoln}
\phi=\phi_0 -\beta t \ ,
\end{equation}
where $\beta\equiv m^2/3\alpha$.
Now, the slow-roll conditions remain satisfied provided that 
\begin{equation}
\phi \gg \frac{\beta}{m} = \frac{\mpl}{\sqrt{12\pi}} \ ,
\end{equation}
and therefore,~(\ref{attractorsoln}) is valid for a time
\begin{equation}
\Delta t \sim \frac{\phi_0}{\beta} \sim \frac{t_P}{\varepsilon^2} \ .
\end{equation}

Why is this important? This is important because~(\ref{attractorsoln})
is an attractor solution! Let's see how this arises. Consider
perturbing~(\ref{attractorsoln}) by writing
\begin{equation}
\phi(t) \rightarrow \phi(t)+\chi(t) \ ,
\end{equation}
where $\chi(t)$ is a ``small" perturbation, substituting in to the
equation of motion and linearizing in $\chi$. One obtains
\begin{equation}
{\ddot \chi}+3\alpha\phi {\dot \chi}=0 \ .
\end{equation}

This equation exhibits two solutions. The first is just a constant,
which just serves to move along the trajectory. The second solution is
a decaying mode with time constant $t_c=(3\alpha\phi)^{-1}$. Since
$t_c \ll \Delta t$, all solutions rapidly decay
to~(\ref{attractorsoln}) - it is an attractor.

\subsection{Solving the Problems of the Standard Cosmology}

Let's stick with the simple model of the last section. It is rather
easy to see that the Einstein equations are solved by
\begin{equation}
a(t)=a(0)\exp\left[\frac{2\pi}{\mpl^2}(\phi_0^2-\phi^2)\right] \ ,
\end{equation}
as we might expect in inflation. The period of time during
which~(\ref{attractorsoln}) is valid ends at $t\sim \Delta t$, at
which
\begin{equation}
a(\Delta t)\sim a(0)\exp\left(\frac{2\pi}{\varepsilon^2}\right) \ ,
\end{equation}
Now, taking a typical value for $m$, for which $\varepsilon <
10^{-4}$, we obtain
\begin{equation}
a(\Delta t) \sim a(0)\times 10^{2.7\times 10^8} \ .
\end{equation}

This has a remarkable consequence. A proper distance $L_P$ at $t=0$
will inflate to a size $10^{10^8}$cm after a time $\Delta t \sim
5\times 10^{-36}$ seconds. It is important at this point to know that
the size of the observable universe today is $H_0^{-1} \sim
and has not had enough 
10^{28}$cm! Therefore, only a small fraction of the original Planck
length comprises today's entire visible universe. Thus, homogeneity over a
patch less than or of order the Planck length at the onset of
inflation is all that is required to solve the horizon problem. Of
course, if we wait sufficiently long we will start to see those
inhomogeneities (originally sub-Planckian) that were inflated
away. However, if inflation lasts long enough (typically about sixty
e-folds or so) then this would not be apparent today.  Similarly,
any unwanted relics are diluted by the tremendous expansion; so long
as the GUT phase transition happens before inflation, monopoles will
have an extremely low density.

Inflation also solves the flatness problem. There are a couple of ways
to see this. The first is to note that, assuming inflation begins, the
curvature term in the Friedmann equation very quickly becomes
irrelevant, since it scales as $a(t)^{-2}$. Of course, after
inflation, when the universe is full of radiation (and later dust) the
curvature term redshifts more slowly than both these components and
will eventually become more important than both of them. However, if
inflation lasts sufficiently long then even today the total energy
density will be so close to unity that we will notice no curvature.

A second way to see this is that, following a similar analysis to that
leading to~(\ref{omegaofa}) leads to the conclusion that 
$\Omega=1$ is an {\it attractor} rather than a repeller, when
the universe is dominated by energy with $w< -1/3$.
Therefore, $\Omega$ is forced to be very close to one by inflation;
if the inflationary period lasts sufficiently long, the density
parameter will not have had time since to stray 
appreciably away from unity.

\subsection{Vacuum Fluctuations and Perturbations}

Recall that the structures - clusters and superclusters of galaxies -
we see on the largest scales in the universe today, and hence the
observed fluctuations in the CMB, form from the gravitational
instability of initial perturbations in the matter density. The origin
of these initial fluctuations is an important question of modern
cosmology.

Inflation provides us with a fascinating solution to this problem - in
a nutshell, quantum fluctuations in the inflaton field during the
inflationary epoch are stretched by inflation and ultimately become
classical fluctuations. Let's sketch how this works.

Since inflation dilutes away all matter fields, soon after its onset
the universe is in a pure vacuum state. If we simplify to the case of
exponential inflation, this vacuum state is described by the
Gibbons-Hawking temperature
\begin{equation}
\label{hawkingtemp}
T_{\rm GH}=\frac{H}{2\pi} \simeq \frac{\sqrt{V}}{\mpl} \ ,
\end{equation}
where we have used the Friedmann equation. Because of this
temperature, the inflaton experiences fluctuations that are the same
for each wavelength $\delta\phi_k=T_{\rm GH}$. Now, these fluctuations
can be related to those in the density by
\begin{equation}
\label{densityflucts}
\delta\rho=\frac{dV}{d\phi} \delta\phi \ .
\end{equation}

Inflation therefore produces density perturbations on every scale.
The amplitude of the perturbations is nearly equal at
each wavenumber, but there will be slight deviations due to the
gradual change in $V$ as the inflaton rolls.  We can
characterize the fluctuations in terms of their spectrum
$A_{\rm S}(k)$, related to the potential via
\be
  A_{\rm S}^2(k) \sim \left.{{V^3}\over{\mpl^6(V')^2}}\right|_{k=aH}\ ,
  \label{scalarspectrum}
\ee
where $k=aH$ indicates that the quantity $V^3/(V')^2$ is to be
evaluated at the moment when the physical scale of the
perturbation $\lambda=a/k$ is equal to the Hubble radius
$H^{-1}$.  Note that the actual normalization of 
(\ref{scalarspectrum}) is convention-dependent, and should drop
out of any physical answer.

The spectrum is given the subscript ``S'' because it describes
scalar fluctuations in the metric.  These are tied to the
energy-momentum distribution, and the density fluctuations
produced by inflation are adiabatic
--- fluctuations in the density of all species
are correlated.  The fluctuations are also Gaussian, in the
sense that the phases of the Fourier modes describing fluctuations
at different scales are uncorrelated.  These aspects of 
inflationary perturbations --- a nearly scale-free spectrum
of adiabatic density fluctuations with a Gaussian distribution ---
are all consistent with current observations of the CMB and 
large-scale structure, and have been confirmed to new precision
by WMAP and other CMB measurements.

It is not only the nearly-massless inflaton that is excited
during inflation, but any nearly-massless particle.  The
other important example is the graviton, which corresponds
to tensor perturbations in the metric (propagating excitations
of the gravitational field).  Tensor fluctuations have a spectrum
\be
  A_{\rm T}^2(k) \sim \left.{{V}\over{\mpl^4}}\right|_{k=aH}\ .
\ee
The existence of tensor perturbations is a crucial prediction
of inflation which may in principle be verifiable through
observations of the polarization of the CMB.  Although CMB
polarization has already been detected \cite{kovac}, this is
only the $E$-mode polarization induced by density perturbations;
the $B$-mode polarization induced by gravitational waves is 
expected to be at a much lower level, and represents a significant
observational challenge for the years to come.

For purposes of understanding observations, it is useful to
parameterize the perturbation spectra in terms of observable
quantities.  We therefore write
\be
  A_{\rm S}^2(k) \propto k^{n_{\rm S}-1}
\ee
and
\be
  A_{\rm T}^2(k) \propto k^{n_{\rm T}}\ ,
\ee
where $n_{\rm S}$ and $n_{\rm T}$ are the ``spectral indices''.  They are
related to the slow-roll parameters of the potential by
\be
  n_{\rm S} = 1 -6\epsilon + 2\eta
\ee
and
\be
  n_{\rm T} = -2\epsilon\ .
\ee
Since the spectral indices are in principle observable, we
can hope through relations such as these to glean some information
about the inflaton potential itself.

Our current knowledge of the amplitude of the perturbations
already gives us important information about the energy scale
of inflation.  Note that the tensor perturbations depend on
$V$ alone (not its derivatives), so observations of tensor
modes yields direct knowledge of the energy scale.
If large-scale CMB anisotropies have an appreciable tensor
component (possible, although unlikely), we can instantly
derive $V_{\rm inflation}\sim (10^{16}$~GeV$)^4$.  
(Here, the value of $V$ being constrained is that which was
responsible for creating the observed fluctuations; namely,
60 $e$-folds before the end of inflation.)  This is 
remarkably reminiscent of the grand unification scale, which
is very encouraging.  Even in the more likely case that the
perturbations observed in the CMB are scalar in nature, we
can still write
\be
  V_{\rm inflation}^{1/4}\sim \epsilon^{1/4}10^{16}~{\rm GeV}\ ,
\ee
where $\epsilon$ is the slow-roll parameter defined in 
(\ref{epsilonslowroll}).
Although we expect $\epsilon$ to be small, the $1/4$ in the
exponent means that the dependence on $\epsilon$ is quite weak;
unless this parameter is extraordinarily tiny, it is very
likely that $V_{\rm inflation}^{1/4}\sim 10^{15}$-$10^{16}$~GeV.

We should note that this discussion has been phrased in terms of
the simplest models of inflation, featuring a single canonical,
slowly-rolling scalar field.  A number of more complex models have
been suggested, allowing for departures from the relations between
the slow-roll parameters and observable quantities; some of these
include hybrid inflation \cite{Linde:1991km,Linde:1993cn,Copeland:1994vg},
inflation with novel kinetic terms \cite{Armendariz-Picon:1999rj},
the curvaton model \cite{Enqvist:2001zp,Lyth:2001nq,Moroi:2001ct},
low-scale models \cite{Randall:fr,Dvali:2003vv},
brane inflation \cite{Dvali:1998pa,Alexander:2001ks,Dvali:2001fw,
Burgess:2001fx,Shiu:2001sy,Choudhury:2003vr,Kachru:2003sx}
and models where perturbations arise from modulated coupling
constants \cite{Bassett:1999cg,Finelli:2000ya,Dvali:2003em,Kofman:2003nx,
Vernizzi:2003vs}.  
This list is necessarily incomplete, and
continued exploration of the varieties of inflationary cosmology
will be a major theme of theoretical cosmology into the foreseeable
future.

\subsection{Reheating and Preheating}
Clearly, one of the great strengths of inflation is its ability to
redshift away all unwanted relics, such as topological
defects. However, inflation is not discerning, and in doing so any
trace of radiation or dust-like matter is similarly redshifted away to
nothing. Thus, at the end of inflation the universe contains nothing
but the inflationary scalar field condensate. How then does that
matter of which we are made arise? How does the hot big bang phase of
the universe commence? How is the universe {\it reheated}?

For a decade or so reheating was thought to be a relatively
well-understood phenomenon. However, over the last 5-10 years, it has
been realized that the story may be quite complicated. We will not
have time to go into details here. Rather, we will sketch the standard
picture and mention briefly the new developments.

Inflation ends when the slow-roll conditions are violated and, in most
models, the field begins to fall towards the minimum of its
potential. Initially, all energy density is in the inflaton, but this
is now damped by two possible terms. First, the expansion of the
universe naturally damps the energy density. More importantly, the
inflaton may decay into other particles, such as radiation or massive
particles, both fermionic and bosonic~\cite{Dolgov:1982th,Abbott:1982hn}. 
To take account of this one
introduces a phenomenological decay term $\Gamma_{\phi}$ into the
scalar field equation. If we focus on fermions only, then a rough
expression for how the energy density evolves is
\begin{equation}
{\dot \rho}_{\phi}+(3H+\Gamma_{\phi})\rho_{\phi}=0 \ .
\end{equation}
The inflaton undergoes damped oscillations and decays into radiation
which equilibrates rapidly at a temperature known as the {\it reheat
temperature} $T_{\rm RH}$.

In the case of bosons however, this ignores the fact that the inflaton
oscillations may give rise to parametric resonance. This is signified
by an extremely rapid decay, yielding a distribution of products that
is far from equilibrium, and only much later settles down to an
equilibrium distribution at energy $T_{\rm RH}$. Such a rapid decay
due to parametric resonance is known as 
{\it preheating}~\cite{Traschen:1990sw,Kofman:1994rk}.

One interesting outcome of preheating is that one can produce
particles with energy far above the ultimate reheat temperature. Thus,
one must beware of producing objects, such as topological relics, that
inflation had rid us of. However, preheating can provide some useful
effects, such as interesting new ways to generate dark matter
candidates and topological relics and to generate the baryon 
asymmetry~\cite{Kofman:1995fi,Tkachev:1995md,Kolb:1996jt,Kasuya:1997ha,Krauss:1999ng,Garcia-Bellido:1999sv}.

\subsection{The Beginnings of Inflation}
If inflation begins, we have seen that it achieves remarkable
results. It can be shown quite generally that the start of inflation
in a previously non-inflating region requires pre-existing homogeneity
over a horizon volume
\cite{Vachaspati:1998dy}. It seems unlikely that classical, causal
physics could cause this and we may interpret this as an indication
that quantum physics may play a role in the process of inflation.

\begin{figure}[htb]
\label{eternal}
\centerline{
\psfig{figure=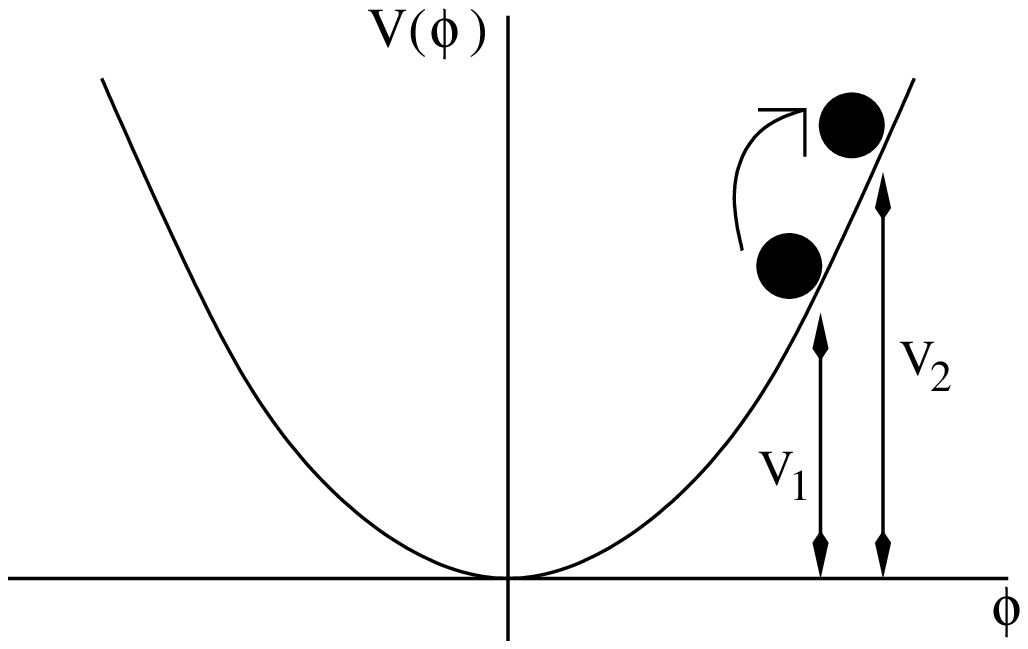,height=2in}}
\caption{}
\end{figure}

Although it can be
shown~\cite{Borde:2001nh,lindepast,steinhardtpast,borde1past,borde2past}
that inflation cannot be past eternal, one consequence of quantum
mechanics is that inflation may be future eternal
\cite{Vilenkin:1983xq,Linde:1986fd,Linde:1995ck,
Vilenkin:1995yd,Linde:1994xx}. To see this,
consider inflation driven by a scalar field rolling down a simple
potential as in figure~(\ref{eternal}).

Imagine that, in a given space-like patch, inflation is occurring with
the scalar field at a potential energy $V_1$. Although the classical
motion of the scalar field is towards the minimum of the potential,
there is a nonzero probability that quantum fluctuations will cause
the inflaton to jump up to a higher value of the potential $V_2 > V_1$
over a sufficiently large part of the inflating patch. In the new
region, the Friedmann equation tells us that the local Hubble
parameter is now larger than in the inflating background
\begin{equation}
H_2 = \sqrt{\frac{V_2}{3\mpl^2}} > \sqrt{\frac{V_1}{3\mpl^2}} =H_1 \ ,
\end{equation}
and requiring the region to be sufficiently large means that its
linear size is at least $H_2^{-1}$, which may be considerably smaller
than the size of the background inflationary horizon.
Thus, inflation occurs in our new region with a larger Hubble
parameter and, because of the exponential nature of inflation, the new
patch very quickly contains a much larger volume of space than the old
one.

There are, of course, parts of space in which the classical motion of
the inflaton to its minimum completes, and those regions drop out of
the inflationary expansion and are free to become regions such of our
own. However, it is interesting to note that, as long as the
probability for inflation to begin at all is not precisely zero, any
region like ours observed at a later time comes from a previously
inflating region with probability essentially unity. Nevertheless, in
this picture, most regions of the universe inflate forever and
therefore this process is known as {\it eternal inflation}.

If correct, eternal inflation provides us with an entirely new global
view of the universe and erases many of our worries about the initial
condition problems of inflation. However, some caution is in order
when considering this model. Eternal inflation relies on the quantum
backreaction of geometry to matter fluctuations over an extended
spatial region. This is something we do not know how to treat fully at
the present time. Perhaps string theory can help us with this.

\section*{Acknowledgements}

We would each like the thank the TASI organizers (Howard Haber,
K.T.~Mahantappa, Juan Maldacena, and Ann Nelson) for inviting us to
lecture and for creating such a vigorous and stimulating environment
for students and lecturers alike.  We would also like to thank the
Kavli Institute for Theoretical Physics for hospitality while these
proceedings were being completed.

The work of SMC is supported in part by U.S. Dept.\ of Energy contract
DE-FG02-90ER-40560, National Science Foundation Grant PHY01-14422
(CfCP) and the David and Lucile Packard Foundation.  The work of MT is
supported in part by the National Science Foundation (NSF) under grant
PHY-0094122. Mark Trodden is a Cottrell Scholar of Research
Corporation.

\end{document}